\def\year{2018}\relax
\documentclass[letterpaper]{article} 

\usepackage{aaai18}  
\usepackage{times}  
\usepackage{helvet}  
\usepackage{courier}  
\usepackage{url}  
\usepackage{graphicx}  
\usepackage{amsmath}
\usepackage{amssymb}
\usepackage{mathtools}

\usepackage{amsthm}
\usepackage[linesnumbered,ruled,vlined]{algorithm2e}

\usepackage{tikz}
\usepackage{tikz-cd}
\usepackage{gnuplot-lua-tikz}

\newcommand{\colora}{0.00,0.00,0.80}
\newcommand{\colorb}{0.70,0.00,0.00}
\newcommand{\colorc}{0.60,0.00,0.60}
\newcommand{\colord}{0.00,0.40,0.50}

\definecolor{xblue}{rgb}{\colora}
\definecolor{xred}{rgb}{\colorb}
\definecolor{xpurple}{rgb}{\colorc}
\definecolor{xgreen}{rgb}{\colord}

\theoremstyle{definition}
\newtheorem{example}{Example}

\begin{document}

\DeclareRobustCommand\model{\mathrel{|}\joinrel\mkern-.5mu\mathrel{-}}
\newcommand{\wrt}{w.r.t.\ }
\newcommand{\var}{\mathit{var}}
\newcommand{\mdef}{\mathrel{:=}}
\newcommand{\apply}[1]{{{\hspace{.1em}|\hspace{.075em}}\lower .25ex\hbox{$#1$}}}
\newcommand{\true}{1}
\newcommand{\false}{0}
\newcommand{\eclause}{\bot}
\newcommand{\sclause}{\top}
\newcommand{\cspace}{\hspace{4pt}}
\newcommand{\aspace}{\hspace{1.8pt}}

\newcommand{\plit}[2]{v_{#1}^{#2}}
\newcommand{\nlit}[2]{\overline v_{#1}^{#2}}
\newcommand{\plita}[2]{\textcolor{xred}{v_{#1}^{#2}}}
\newcommand{\nlita}[2]{\textcolor{xred}{\overline v_{#1}^{#2}}}
\newcommand{\plitb}[2]{\textcolor{xblue}{v_{#1}^{#2}}}
\newcommand{\nlitb}[2]{\textcolor{xblue}{\overline v_{#1}^{#2}}}

\title{Schur Number Five}
\author{Marijn J.H. Heule\\
Computer Science Department\\The University of Texas at Austin\\
2317 Speedway, M/S D9500\\
Austin, Texas 78712-0233
}
\maketitle
\begin{abstract}
We present the solution of a century-old problem known as \emph{Schur Number Five}:
What is the largest (natural) number $n$ such that there exists a five-coloring of the positive numbers up to $n$
without a monochromatic solution of the equation $a + b = c$? 
We obtained the solution, $n = 160$, by encoding the problem into propositional logic
and applying massively parallel satisfiability solving techniques on the resulting formula. 
We constructed and validated a proof of the solution to increase trust in the correctness 
of the multi-CPU-year computations. The proof is two petabytes in
size and was certified using a formally verified proof checker, demonstrating that 
any result by satisfiability solvers---no matter how large---can now be validated using highly trustworthy systems.
\end{abstract}

\section{Introduction}

In the beginning of the $20^\mathrm{th}$ century, Issai Schur studied whether every coloring
of the positive (natural) numbers with finitely many colors results in monochromatic
solutions of the equation $a + b = c$.
This work gave rise to the concept of so-called \emph{Schur numbers}:
Schur number $k$, denoted by $S(k)$, is defined as the
largest number $n$ for which there exists a $k$-coloring of the positive numbers up to $n$ with no
monochromatic solution of $a + b = c$.\footnote{An alternative definition used
in the literature picks the smallest $n$ s.t. all $k$-colorings of 1 to $n$ result in a monochromatic solution.
The values of $S(k)$ differ by one, depending on the definition.}
For example, $S(2) = 4$: Assume we use the two colors red and blue.
If we color $1$ with red, we have to color $2$ with blue due to $1 + 1 = 2$.
This forces us to color $4$ with red because of $2 + 2 = 4$. After this, $3$
must become blue due to $1 + 3 = 4$. 
But then, no matter if we color 5 with red or blue, we end up with a
monochromatic solution of $1 + 4 = 5$ or $2 + 3 = 5$.


Although Schur's Theorem states that $S(k)$ is finite
for any finite value of $k$~\cite{Sch17},  
determining the exact values of $S(k)$ is 
an open
problem in elementary number theory~\cite{Guy}.
In fact, only the values $S(1) = 1$, $S(2) = 4$,
$S(3) = 13$, and $S(4) = 44$ have been known so far~\cite{S44}. 
We came up with a highly optimized automated-reasoning method for showing that $S(5) = 160$.

To obtain this solution, we first encoded the Schur Number Five problem into propositional logic and then applied 
satisfiability (SAT) solving techniques to solve the resulting formula. 
This approach has been successful in recent years, leading to the solution of 
hard open problems such as the problem of determining the sixth van der Waerden number~\cite{KP08}, 
the Erd\H{o}s discrepancy problem~\cite{Konev:2015}, and the Pythagorean triples problem~\cite{ptn}.
Trying to solve a SAT encoding for Schur Number Five with off-the-shelf SAT solving tools 
turned out to be a hopeless endeavor. 
We therefore came up with a dedicated approach, which is intended to be applicable to related problems as well.
We modified existing tools to efficiently solve our encoding. 
Still, even with our optimized approach, the total computational effort to solve the problem was over $14$ CPU years.

If it takes a computer several CPU years to solve a problem, it is only natural to question the correctness of
the supposed solution.
To deal with this issue, we automatically constructed a proof of the propositional formula that encodes the main
statement.
The size of this proof is more than two petabytes, making it about ten times larger than
``the largest math proof ever''~\cite{Lamb}.
Despite its tremendous size, we were able to verify the correctness of the proof with a 
formally verified proof checker. 
Due to recent progress in proof validation~\cite{Cruz-FilipeMS17}, checking the correctness of such proofs is now
nearly as efficient as the actual construction of the proofs by a SAT solver~\cite{Cruz-Filipe2017,Lammich2017}.
In our case, the time spent on proof checking was a little more than $36$ CPU years. 


The main contributions of this paper are as follows:
\begin{itemize}
\item We constructed a propositional formula that is satisfiable if and only if $S(5) \geq 161$. 
Our proof of unsatisfiability for this formula is over two petabytes in size.
\item We certified the proof using a program formally verified by 
ACL2~\cite{ACL2}, thereby providing high confidence in the correctness of our result.
\item We enumerated all $2\,447\,113\,088$ five-colorings of the numbers 1 to 160 without a monochromatic $a + b = c$. 
\item We designed a decision heuristic that allows solving Schur number problems efficiently and enables linear-time speedups even when using thousands of CPUs.
\item We developed an efficient hardness predictor for partitioning a hard problem into millions of easy subproblems. 
\end{itemize}

\section{Schur Numbers and Variants}

Schur number $k$, denoted by $S(k)$, is defined as the largest (natural) number $n$ such that there
exists a $k$-coloring of the numbers $1$ to $n$ without a monochromatic solution 
of the equation $a+b=c$  with $1 \leq a,b,c\leq n$. 
The first Schur numbers $S(1) = 1$, $S(2) = 4$, and $S(3) = 13$ can be determined manually while 
$S(4) = 44$ was computed decades ago~\cite{S44}. The best known lower bounds for 
higher Schur numbers
are: $S(5) \geq 160$~\cite{Exoo}, $S(6) \geq 536$, and $S(7) \geq 1680$~\cite{Palin}.
We prove that $S(5) = 160$.

The early upper bounds $S(k) \leq \lfloor k!e \rfloor$~\cite{Sch17} have later
been improved to $S(k) \leq \lfloor k!(e- \frac{1}{24}) \rfloor$~\cite{extension}.
Upper bounds on $S(k)$ can also be obtained via the connection to 
the Ramsey numbers $R_k(3)$, which denote the smallest $n$ such that
any $k$-coloring of the edges of the fully connected graph on $n$
vertices yields a monochromatic triangle: $S(k) \leq R_k(3) - 2$~\cite{Sch17}.
The first three numbers $R_k(3)$ are known:
$R_1(3) = 3$, $R_2(3) = 6$, and $R_3(3) = 17$~\cite{RamseyR333}.

Several variants of Schur numbers have been proposed.
The oldest variant, known as \emph{weak} Schur number $k$ and denoted by $W\!S(k)$, requires $a$ to be smaller than $b$, 
thus weakening $1 \leq a,b,c\leq n$ to $1 \leq a < b < c\leq n$~\cite{extension}. Hence, $W\!S(k) \geq S(k)$. Only the four smallest weak Schur numbers
are known:  $W\!S(1) = 2$, $W\!S(2) = 8$, $W\!S(3) = 23$, and $W\!S(4) = 66$~\cite{WS4}.

Another variant is the \emph{modular} Schur number $k$, denoted by $S_\mathrm{mod} (k)$,
asking for the largest $n$ such that a 
$k$-coloring of the numbers $1$ to $n$ exists without a monochromatic solution 
of the equation $a+b \equiv c \pmod{n+1}$ with $1 \leq a,b,c\leq n$ \cite{modular}. 
This variant is stronger than the classical notion, hence $S(k) \geq S_\mathrm{mod} (k)$.
However, for all known Schur numbers it holds that $S(k) = S_\mathrm{mod} (k)$
and this equality is conjectured to hold in general~\cite{modular}.
Our result implies the equality for $k = 5$. 

An even stronger variant is the \emph{palindromic} Schur number $k$,
denoted by $S_\mathrm{pd} (k)$, for which the numbers $i$ and $n+1-i$ with $1 \leq i \leq n/2$  have
the same color---except in case $2i = n+1-i$. 
This variant is also known 
as \emph{symmetric sum-free sets}~\cite{Palin} and is mainly used to determine lower bounds for the weaker variants.  
We have $S_\mathrm{mod} (k) \geq S_\mathrm{pd} (k)$ in general. However, the numbers are equal for the known values.
This is a new result for $k=5$.

The big question is whether $S(k) = S_\mathrm{mod} (k) = S_\mathrm{pd} (k)$ for any $k$. 
We can probably answer this question only if the answer is no. Already showing this equality
for $k=6$ is expected to be extremely challenging.

\section{Technical Background}
\label{sec:prelim}

Below we present the most important background concepts related to the more technical part of this paper.

\paragraph{Propositional logic.}
We consider propositional formulas in \emph{conjunctive normal form} (CNF), 
which are defined as follows. 
A \emph{literal} is either a variable $v$ (a \emph{positive literal}) 
or the negation $\overline v$ of a variable~$v$ (a \emph{negative literal}). 
The \emph{complementary literal} $\overline l$ of a literal $l$ is defined as 
$\overline l = \overline v$ if $l = v$ and $\overline l = v$ if $l = \overline v$.
A \emph{clause} is a disjunction of literals.
A \emph{formula} is a conjunction of clauses.
For a literal, clause, or formula $F$, $\var(F)$ denotes the variables in $F$. 
For convenience, we treat $\var(F)$ as a variable if $F$ is a literal, and as a set of variables otherwise.

\paragraph{Satisfiability.} An \emph{assignment} is a
function from a set of variables to the truth values 
\true{}~(\emph{true}) and \false{} (\emph{false}).
An assignment is \emph{total} \wrt a formula if it assigns a truth value to all
variables occurring in the formula; otherwise it is $\emph{partial}$.
A literal $l$ is \emph{satisfied} (\emph{falsified}) by an assignment $\alpha$ if 
$l$ is positive and \mbox{$\alpha(\var(l)) = \true$} ($\alpha(\var(l)) = \false$, resp.) 
or if it is negative and $\alpha(\var(l)) = \false$ ($\alpha(\var(l)) = \true$, resp.).
We also denote with $\alpha$ the conjuction of literals that are satisfied by that assignment;
such a conjunction is called a \emph{cube}.
A clause is satisfied by an assignment $\alpha$ if it contains a literal that is satisfied by~$\alpha$.
Finally, a formula is satisfied by an assignment $\alpha$ if all its clauses are satisfied by $\alpha$.
A formula is \emph{satisfiable} if there exists an assignment that satisfies it; otherwise it is {\emph{unsatisfiable}.
A formula $F$ \emph{entails} a formula $G$, denoted by $F \models G$, if every assignment that satisfies
$F$ also satisfies $G$. $F$ \emph{weakly entails} $G$, denoted by $F \models_{\!\!\!\!\mathrm{_w}} G$,
if satisfiability of $F$ implies satisfiability of $G$.


\paragraph{Proofs of Unsatisfiability.}

It is easy to check that an alleged satisfying assignment is valid. However, a certificate that a formula 
has no solution (i.e., is unsatisfiable) can be huge and costly to validate. 
We produce proofs of unsatisfiability in the DRAT proof system~\cite{inprocessing}, which is the standard
in state-of-the-art SAT solving.
Given a formula $F$, a DRAT proof of unsatisfiability is a 
sequence $C_1, \dots, C_m$ of clauses where $C_m$ is the empty clause $\eclause$.
For every clause $C_i$, it must hold that $C_i$ is a \emph{resolution asymmetric tautology} (RAT)
with respect to $F \cup \{C_1, \dots, C_{i-1}\}$. The addition of a RAT to a formula
preserves satisfiability and since the empty clause is trivially unsatisfiable, a DRAT proof
witnesses the unsatisfiability of the original formula~$F$.
DRAT also allows the deletion of clauses from a formula to improve the performance of proof validation.
Note that clause deletion preserves satisfiability.

\newcommand{\aw}{xred}\newcommand{\ax}{0.75}\newcommand{\ay}{1.00}\newcommand{\az}{0.875}
\newcommand{\bw}{xblue}\newcommand{\bx}{0.50}\newcommand{\by}{0.75}\newcommand{\bz}{0.625}
\newcommand{\cw}{xpurple}\newcommand{\cx}{0.25}\newcommand{\cy}{0.50}\newcommand{\cz}{0.375}
\newcommand{\dw}{xgreen}\newcommand{\dx}{0.00}\newcommand{\dy}{0.25}\newcommand{\dz}{0.125}
\newcommand{\ew}{black}\newcommand{\ex}{-0.25}\newcommand{\ey}{0.00}\newcommand{\ez}{-0.125}

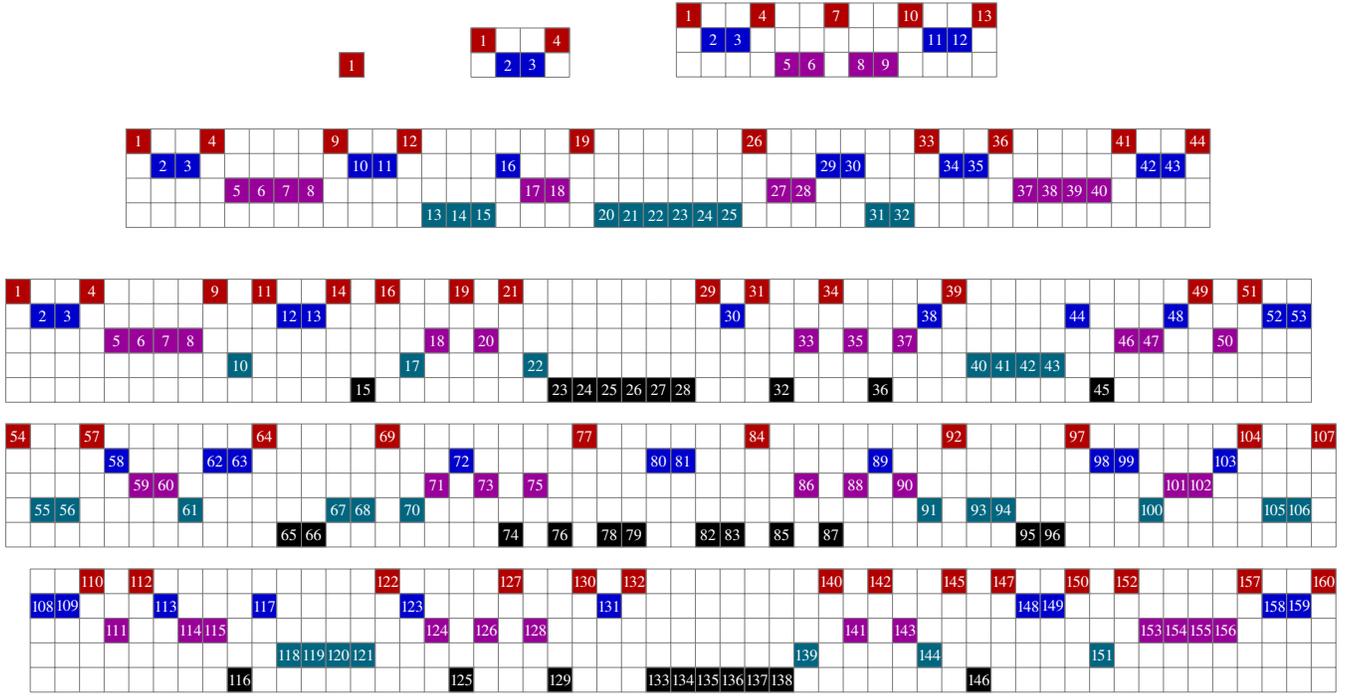
\begin{figure*}
\centering
\begin{tikzpicture}[scale=1.31]
\draw[step=0.25cm,color=gray] (0,0) grid (0.25,0.25);
\filldraw[fill=xred, draw=gray] (0,0) rectangle (0.25,0.25); \node at (0.125,0.125) {\tiny \textcolor{white}{1}};
\end{tikzpicture}
~~~~~~~~~~~~~~
\begin{tikzpicture}[scale=1.31]
\draw[step=0.25cm,color=gray] (0,0) grid (1,0.5);
\filldraw[fill=xred, draw=gray] (0,0.25) rectangle (0.25,0.5); \node at (0.125,0.375) {\tiny \textcolor{white}{1}};
\filldraw[fill=xblue, draw=gray] (0.25,0.0) rectangle (0.50,0.25); \node at (0.375,0.125) {\tiny \textcolor{white}{2}};
\filldraw[fill=xblue, draw=gray] (0.5,0.0) rectangle (0.75,0.25); \node at (0.625,0.125) {\tiny \textcolor{white}{3}};
\filldraw[fill=xred, draw=gray] (0.75,0.25) rectangle (1.00,0.5); \node at (0.875,0.375) {\tiny \textcolor{white}{4}};
\end{tikzpicture}
~~~~~~~~~~~~~~
\begin{tikzpicture}[scale=1.31]
\draw[step=0.25cm,color=gray] (0,0) grid (3.25,0.75);
\filldraw[fill=xred, draw=gray] (0,0.50) rectangle (0.25,0.75); \node at (0.125,0.625) {\tiny \textcolor{white}{1}};
\filldraw[fill=xblue, draw=gray] (0.25,0.25) rectangle (0.50,0.50); \node at (0.375,0.375) {\tiny \textcolor{white}{2}};
\filldraw[fill=xblue, draw=gray] (0.5,0.25) rectangle (0.75,0.50); \node at (0.625,0.375) {\tiny \textcolor{white}{3}};
\filldraw[fill=xred, draw=gray] (0.75,0.50) rectangle (1.00,0.75); \node at (0.875,0.625) {\tiny \textcolor{white}{4}};
\filldraw[fill=xpurple, draw=gray] (1.00,0.00) rectangle (1.25,0.25); \node at (1.125,0.125) {\tiny \textcolor{white}{5}};
\filldraw[fill=xpurple, draw=gray] (1.25,0.00) rectangle (1.50,0.25); \node at (1.375,0.125) {\tiny \textcolor{white}{6}};
\filldraw[fill=xred, draw=gray] (1.50,0.50) rectangle (1.75,0.75); \node at (1.625,0.625) {\tiny \textcolor{white}{7}};
\filldraw[fill=xpurple, draw=gray] (1.75,0.00) rectangle (2.00,0.25); \node at (1.875,0.125) {\tiny \textcolor{white}{8}};
\filldraw[fill=xpurple, draw=gray] (2.00,0.00) rectangle (2.25,0.25); \node at (2.125,0.125) {\tiny \textcolor{white}{9}};
\filldraw[fill=xred, draw=gray] (2.25,0.50) rectangle (2.50,0.75); \node at (2.375,0.625) {\tiny \textcolor{white}{10}};
\filldraw[fill=xblue, draw=gray] (2.50,0.25) rectangle (2.75,0.50); \node at (2.625,0.375) {\tiny \textcolor{white}{11}};
\filldraw[fill=xblue, draw=gray] (2.75,0.25) rectangle (3.00,0.50); \node at (2.875,0.375) {\tiny \textcolor{white}{12}};
\filldraw[fill=xred, draw=gray] (3.00,0.50) rectangle (3.25,0.75); \node at (3.125,0.625) {\tiny \textcolor{white}{13}};
\end{tikzpicture}\\[0.6cm]
\begin{tikzpicture}[scale=1.31]
\draw[step=0.25cm,color=gray] (0,0) grid (11,1);
\filldraw[fill=xred, draw=gray] (0,0.75) rectangle (0.25,1.00); \node at (0.125,0.875) {\tiny \textcolor{white}{1}};
\filldraw[fill=xblue, draw=gray] (0.25,0.50) rectangle (0.50,0.75); \node at (0.375,0.625) {\tiny \textcolor{white}{2}};
\filldraw[fill=xblue, draw=gray] (0.5,0.50) rectangle (0.75,0.75); \node at (0.625,0.625) {\tiny \textcolor{white}{3}};
\filldraw[fill=xred, draw=gray] (0.75,0.75) rectangle (1.00,1.00); \node at (0.875,0.875) {\tiny \textcolor{white}{4}};
\filldraw[fill=xpurple, draw=gray] (1.00,0.25) rectangle (1.25,0.50); \node at (1.125,0.375) {\tiny \textcolor{white}{5}};
\filldraw[fill=xpurple, draw=gray] (1.25,0.25) rectangle (1.50,0.50); \node at (1.375,0.375) {\tiny \textcolor{white}{6}};
\filldraw[fill=xpurple, draw=gray] (1.50,0.25) rectangle (1.75,0.50); \node at (1.625,0.375) {\tiny \textcolor{white}{7}};
\filldraw[fill=xpurple, draw=gray] (1.75,0.25) rectangle (2.00,0.50); \node at (1.875,0.375) {\tiny \textcolor{white}{8}};
\filldraw[fill=xred, draw=gray] (2.00,0.75) rectangle (2.25,1.00); \node at (2.125,0.875) {\tiny \textcolor{white}{9}};
\filldraw[fill=xblue, draw=gray] (2.25,0.50) rectangle (2.50,0.75); \node at (2.375,0.625) {\tiny \textcolor{white}{10}};
\filldraw[fill=xblue, draw=gray] (2.50,0.50) rectangle (2.75,0.75); \node at (2.625,0.625) {\tiny \textcolor{white}{11}};
\filldraw[fill=xred, draw=gray] (2.75,0.75) rectangle (3.00,1.00); \node at (2.875,0.875) {\tiny \textcolor{white}{12}};
\filldraw[fill=\dw, draw=gray] (3.00,0.00) rectangle (3.25,0.25); \node at (3.125,0.125) {\tiny \textcolor{white}{13}};
\filldraw[fill=\dw, draw=gray] (3.25,0.00) rectangle (3.50,0.25); \node at (3.375,0.125) {\tiny \textcolor{white}{14}};
\filldraw[fill=\dw, draw=gray] (3.50,0.00) rectangle (3.75,0.25); \node at (3.625,0.125) {\tiny \textcolor{white}{15}};
\filldraw[fill=xblue, draw=gray] (3.75,0.50) rectangle (4.00,0.75); \node at (3.875,0.625) {\tiny \textcolor{white}{16}};
\filldraw[fill=xpurple, draw=gray] (4.00,0.25) rectangle (4.25,0.50); \node at (4.125,0.375) {\tiny \textcolor{white}{17}};
\filldraw[fill=xpurple, draw=gray] (4.25,0.25) rectangle (4.50,0.50); \node at (4.375,0.375) {\tiny \textcolor{white}{18}};
\filldraw[fill=xred, draw=gray] (4.50,0.75) rectangle (4.75,1.00); \node at (4.625,0.875) {\tiny \textcolor{white}{19}};
\filldraw[fill=\dw, draw=gray] (4.75,0.00) rectangle (5.00,0.25); \node at (4.875,0.125) {\tiny \textcolor{white}{20}};
\filldraw[fill=\dw, draw=gray] (5.00,0.00) rectangle (5.25,0.25); \node at (5.125,0.125) {\tiny \textcolor{white}{21}};
\filldraw[fill=\dw, draw=gray] (5.25,0.00) rectangle (5.50,0.25); \node at (5.375,0.125) {\tiny \textcolor{white}{22}};
\filldraw[fill=\dw, draw=gray] (5.50,0.00) rectangle (5.75,0.25); \node at (5.625,0.125) {\tiny \textcolor{white}{23}};
\filldraw[fill=\dw, draw=gray] (5.75,0.00) rectangle (6.00,0.25); \node at (5.875,0.125) {\tiny \textcolor{white}{24}};
\filldraw[fill=\dw, draw=gray] (6.00,0.00) rectangle (6.25,0.25); \node at (6.125,0.125) {\tiny \textcolor{white}{25}};
\filldraw[fill=xred, draw=gray] (6.25,0.75) rectangle (6.50,1.00); \node at (6.375,0.875) {\tiny \textcolor{white}{26}};
\filldraw[fill=xpurple, draw=gray] (6.50,0.25) rectangle (6.75,0.50); \node at (6.625,0.375) {\tiny \textcolor{white}{27}};
\filldraw[fill=xpurple, draw=gray] (6.75,0.25) rectangle (7.00,0.50); \node at (6.875,0.375) {\tiny \textcolor{white}{28}};
\filldraw[fill=xblue, draw=gray] (7.00,0.50) rectangle (7.25,0.75); \node at (7.125,0.625) {\tiny \textcolor{white}{29}};
\filldraw[fill=xblue, draw=gray] (7.25,0.50) rectangle (7.50,0.75); \node at (7.375,0.625) {\tiny \textcolor{white}{30}};
\filldraw[fill=\dw, draw=gray] (7.50,0.00) rectangle (7.75,0.25); \node at (7.625,0.125) {\tiny \textcolor{white}{31}};
\filldraw[fill=\dw, draw=gray] (7.75,0.00) rectangle (8.00,0.25); \node at (7.875,0.125) {\tiny \textcolor{white}{32}};
\filldraw[fill=xred, draw=gray] (8.00,0.75) rectangle (8.25,1.00); \node at (8.125,0.875) {\tiny \textcolor{white}{33}};
\filldraw[fill=xblue, draw=gray] (8.25,0.50) rectangle (8.50,0.75); \node at (8.375,0.625) {\tiny \textcolor{white}{34}};
\filldraw[fill=xblue, draw=gray] (8.50,0.50) rectangle (8.75,0.75); \node at (8.625,0.625) {\tiny \textcolor{white}{35}};
\filldraw[fill=xred, draw=gray] (8.75,0.75) rectangle (9.00,1.00); \node at (8.875,0.875) {\tiny \textcolor{white}{36}};
\filldraw[fill=xpurple, draw=gray] (9.00,0.25) rectangle (9.25,0.50); \node at (9.125,0.375) {\tiny \textcolor{white}{37}};
\filldraw[fill=xpurple, draw=gray] (9.25,0.25) rectangle (9.50,0.50); \node at (9.375,0.375) {\tiny \textcolor{white}{38}};
\filldraw[fill=xpurple, draw=gray] (9.50,0.25) rectangle (9.75,0.50); \node at (9.625,0.375) {\tiny \textcolor{white}{39}};
\filldraw[fill=xpurple, draw=gray] (9.75,0.25) rectangle (10.00,0.50); \node at (9.875,0.375) {\tiny \textcolor{white}{40}};
\filldraw[fill=xred, draw=gray] (10.00,0.75) rectangle (10.25,1.00); \node at (10.125,0.875) {\tiny \textcolor{white}{41}};
\filldraw[fill=xblue, draw=gray] (10.25,0.50) rectangle (10.50,0.75); \node at (10.375,0.625) {\tiny \textcolor{white}{42}};
\filldraw[fill=xblue, draw=gray] (10.50,0.50) rectangle (10.75,0.75); \node at (10.625,0.625) {\tiny \textcolor{white}{43}};
\filldraw[fill=xred, draw=gray] (10.75,0.75) rectangle (11.00,1.00); \node at (10.875,0.875) {\tiny \textcolor{white}{44}};
\end{tikzpicture}\\[0.6cm]

\begin{tikzpicture}[scale=1.31]
\draw[step=0.25cm,color=white] (0,-0.25) grid (13.48,1);
\draw[step=0.25cm,color=gray] (0,-0.25) grid (13.25,1);
\filldraw[fill=\aw, draw=gray] (0,\ax) rectangle (0.25,\ay); \node at (0.125,\az) {\tiny \textcolor{white}{1}};  
\filldraw[fill=\bw, draw=gray] (0.25,\bx) rectangle (0.50,\by); \node at (0.375,\bz) {\tiny \textcolor{white}{2}}; 
\filldraw[fill=\bw, draw=gray] (0.5,\bx) rectangle (0.75,\by); \node at (0.625,\bz) {\tiny \textcolor{white}{3}}; 
\filldraw[fill=\aw, draw=gray] (0.75,\ax) rectangle (1.00,\ay); \node at (0.875,\az) {\tiny \textcolor{white}{4}}; 
\filldraw[fill=\cw, draw=gray] (1.00,\cx) rectangle (1.25,\cy); \node at (1.125,\cz) {\tiny \textcolor{white}{5}}; 
\filldraw[fill=\cw, draw=gray] (1.25,\cx) rectangle (1.50,\cy); \node at (1.375,\cz) {\tiny \textcolor{white}{6}}; 
\filldraw[fill=\cw, draw=gray] (1.50,\cx) rectangle (1.75,\cy); \node at (1.625,\cz) {\tiny \textcolor{white}{7}}; 
\filldraw[fill=\cw, draw=gray] (1.75,\cx) rectangle (2.00,\cy); \node at (1.875,\cz) {\tiny \textcolor{white}{8}}; 
\filldraw[fill=\aw, draw=gray] (2.00,\ax) rectangle (2.25,\ay); \node at (2.125,\az) {\tiny \textcolor{white}{9}}; 
\filldraw[fill=\dw, draw=gray] (2.25,\dx) rectangle (2.50,\dy); \node at (2.375,\dz) {\tiny \textcolor{white}{10}}; 
\filldraw[fill=\aw, draw=gray] (2.50,\ax) rectangle (2.75,\ay); \node at (2.625,\az) {\tiny \textcolor{white}{11}}; 
\filldraw[fill=\bw, draw=gray] (2.75,\bx) rectangle (3.00,\by); \node at (2.875,\bz) {\tiny \textcolor{white}{12}}; 
\filldraw[fill=\bw, draw=gray] (3.00,\bx) rectangle (3.25,\by); \node at (3.125,\bz) {\tiny \textcolor{white}{13}}; 
\filldraw[fill=\aw, draw=gray] (3.25,\ax) rectangle (3.50,\ay); \node at (3.375,\az) {\tiny \textcolor{white}{14}}; 
\filldraw[fill=\ew, draw=gray] (3.50,\ex) rectangle (3.75,\ey); \node at (3.625,\ez) {\tiny \textcolor{white}{15}}; 
\filldraw[fill=\aw, draw=gray] (3.75,\ax) rectangle (4.00,\ay); \node at (3.875,\az) {\tiny \textcolor{white}{16}}; 
\filldraw[fill=\dw, draw=gray] (4.00,\dx) rectangle (4.25,\dy); \node at (4.125,\dz) {\tiny \textcolor{white}{17}}; 
\filldraw[fill=\cw, draw=gray] (4.25,\cx) rectangle (4.50,\cy); \node at (4.375,\cz) {\tiny \textcolor{white}{18}}; 
\filldraw[fill=\aw, draw=gray] (4.50,\ax) rectangle (4.75,\ay); \node at (4.625,\az) {\tiny \textcolor{white}{19}}; 
\filldraw[fill=\cw, draw=gray] (4.75,\cx) rectangle (5.00,\cy); \node at (4.875,\cz) {\tiny \textcolor{white}{20}}; 
\filldraw[fill=\aw, draw=gray] (5.00,\ax) rectangle (5.25,\ay); \node at (5.125,\az) {\tiny \textcolor{white}{21}}; 
\filldraw[fill=\dw, draw=gray] (5.25,\dx) rectangle (5.50,\dy); \node at (5.375,\dz) {\tiny \textcolor{white}{22}}; 
\filldraw[fill=\ew, draw=gray] (5.50,\ex) rectangle (5.75,\ey); \node at (5.625,\ez) {\tiny \textcolor{white}{23}}; 
\filldraw[fill=\ew, draw=gray] (5.75,\ex) rectangle (6.00,\ey); \node at (5.875,\ez) {\tiny \textcolor{white}{24}}; 
\filldraw[fill=\ew, draw=gray] (6.00,\ex) rectangle (6.25,\ey); \node at (6.125,\ez) {\tiny \textcolor{white}{25}}; 
\filldraw[fill=\ew, draw=gray] (6.25,\ex) rectangle (6.50,\ey); \node at (6.375,\ez) {\tiny \textcolor{white}{26}}; 
\filldraw[fill=\ew, draw=gray] (6.50,\ex) rectangle (6.75,\ey); \node at (6.625,\ez) {\tiny \textcolor{white}{27}}; 
\filldraw[fill=\ew, draw=gray] (6.75,\ex) rectangle (7.00,\ey); \node at (6.875,\ez) {\tiny \textcolor{white}{28}}; 
\filldraw[fill=\aw, draw=gray] (7.00,\ax) rectangle (7.25,\ay); \node at (7.125,\az) {\tiny \textcolor{white}{29}}; 
\filldraw[fill=\bw, draw=gray] (7.25,\bx) rectangle (7.50,\by); \node at (7.375,\bz) {\tiny \textcolor{white}{30}}; 
\filldraw[fill=\aw, draw=gray] (7.50,\ax) rectangle (7.75,\ay); \node at (7.625,\az) {\tiny \textcolor{white}{31}}; 
\filldraw[fill=\ew, draw=gray] (7.75,\ex) rectangle (8.00,\ey); \node at (7.875,\ez) {\tiny \textcolor{white}{32}}; 
\filldraw[fill=\cw, draw=gray] (8.00,\cx) rectangle (8.25,\cy); \node at (8.125,\cz) {\tiny \textcolor{white}{33}}; 
\filldraw[fill=\aw, draw=gray] (8.25,\ax) rectangle (8.50,\ay); \node at (8.375,\az) {\tiny \textcolor{white}{34}}; 
\filldraw[fill=\cw, draw=gray] (8.50,\cx) rectangle (8.75,\cy); \node at (8.625,\cz) {\tiny \textcolor{white}{35}}; 
\filldraw[fill=\ew, draw=gray] (8.75,\ex) rectangle (9.00,\ey); \node at (8.875,\ez) {\tiny \textcolor{white}{36}}; 
\filldraw[fill=\cw, draw=gray] (9.00,\cx) rectangle (9.25,\cy); \node at (9.125,\cz) {\tiny \textcolor{white}{37}}; 
\filldraw[fill=\bw, draw=gray] (9.25,\bx) rectangle (9.50,\by); \node at (9.375,\bz) {\tiny \textcolor{white}{38}}; 
\filldraw[fill=\aw, draw=gray] (9.50,\ax) rectangle (9.75,\ay); \node at (9.625,\az) {\tiny \textcolor{white}{39}}; 
\filldraw[fill=\dw, draw=gray] (9.75,\dx) rectangle (10.00,\dy); \node at (9.875,\dz) {\tiny \textcolor{white}{40}}; 
\filldraw[fill=\dw, draw=gray] (10.00,\dx) rectangle (10.25,\dy); \node at (10.125,\dz) {\tiny \textcolor{white}{41}}; 
\filldraw[fill=\dw, draw=gray] (10.25,\dx) rectangle (10.50,\dy); \node at (10.375,\dz) {\tiny \textcolor{white}{42}}; 
\filldraw[fill=\dw, draw=gray] (10.50,\dx) rectangle (10.75,\dy); \node at (10.625,\dz) {\tiny \textcolor{white}{43}}; 
\filldraw[fill=\bw, draw=gray] (10.75,\bx) rectangle (11.00,\by); \node at (10.875,\bz) {\tiny \textcolor{white}{44}};  
\filldraw[fill=\ew, draw=gray] (11.00,\ex) rectangle (11.25,\ey); \node at (11.125,\ez) {\tiny \textcolor{white}{45}}; 
\filldraw[fill=\cw, draw=gray] (11.25,\cx) rectangle (11.50,\cy); \node at (11.375,\cz) {\tiny \textcolor{white}{46}}; 
\filldraw[fill=\cw, draw=gray] (11.50,\cx) rectangle (11.75,\cy); \node at (11.625,\cz) {\tiny \textcolor{white}{47}}; 
\filldraw[fill=\bw, draw=gray] (11.75,\bx) rectangle (12.00,\by); \node at (11.875,\bz) {\tiny \textcolor{white}{48}}; 
\filldraw[fill=\aw, draw=gray] (12.00,\ax) rectangle (12.25,\ay); \node at (12.125,\az) {\tiny \textcolor{white}{49}}; 
\filldraw[fill=\cw, draw=gray] (12.25,\cx) rectangle (12.50,\cy); \node at (12.375,\cz) {\tiny \textcolor{white}{50}}; 
\filldraw[fill=\aw, draw=gray] (12.50,\ax) rectangle (12.75,\ay); \node at (12.625,\az) {\tiny \textcolor{white}{51}}; 
\filldraw[fill=\bw, draw=gray] (12.75,\bx) rectangle (13.00,\by); \node at (12.875,\bz) {\tiny \textcolor{white}{52}}; 
\filldraw[fill=\bw, draw=gray] (13.00,\bx) rectangle (13.25,\by); \node at (13.125,\bz) {\tiny \textcolor{white}{53}}; 
\end{tikzpicture}\\[0.2cm]
\begin{tikzpicture}[scale=1.31]
\draw[step=0.25cm,color=gray] (0,-0.25) grid (13.5,1);
\filldraw[fill=\aw, draw=gray] (0,\ax) rectangle (0.25,\ay); \node at (0.125,\az) {\tiny \textcolor{white}{54}}; 
\filldraw[fill=\dw, draw=gray] (0.25,\dx) rectangle (0.50,\dy); \node at (0.375,\dz) {\tiny \textcolor{white}{55}}; 
\filldraw[fill=\dw, draw=gray] (0.5,\dx) rectangle (0.75,\dy); \node at (0.625,\dz) {\tiny \textcolor{white}{56}}; 
\filldraw[fill=\aw, draw=gray] (0.75,\ax) rectangle (1.00,\ay); \node at (0.875,\az) {\tiny \textcolor{white}{57}}; 
\filldraw[fill=\bw, draw=gray] (1.00,\bx) rectangle (1.25,\by); \node at (1.125,\bz) {\tiny \textcolor{white}{58}}; 
\filldraw[fill=\cw, draw=gray] (1.25,\cx) rectangle (1.50,\cy); \node at (1.375,\cz) {\tiny \textcolor{white}{59}}; 
\filldraw[fill=\cw, draw=gray] (1.50,\cx) rectangle (1.75,\cy); \node at (1.625,\cz) {\tiny \textcolor{white}{60}}; 
\filldraw[fill=\dw, draw=gray] (1.75,\dx) rectangle (2.00,\dy); \node at (1.875,\dz) {\tiny \textcolor{white}{61}}; 
\filldraw[fill=\bw, draw=gray] (2.00,\bx) rectangle (2.25,\by); \node at (2.125,\bz) {\tiny \textcolor{white}{62}}; 
\filldraw[fill=\bw, draw=gray] (2.25,\bx) rectangle (2.50,\by); \node at (2.375,\bz) {\tiny \textcolor{white}{63}}; 
\filldraw[fill=\aw, draw=gray] (2.50,\ax) rectangle (2.75,\ay); \node at (2.625,\az) {\tiny \textcolor{white}{64}}; 
\filldraw[fill=\ew, draw=gray] (2.75,\ex) rectangle (3.00,\ey); \node at (2.875,\ez) {\tiny \textcolor{white}{65}}; 
\filldraw[fill=\ew, draw=gray] (3.00,\ex) rectangle (3.25,\ey); \node at (3.125,\ez) {\tiny \textcolor{white}{66}}; 
\filldraw[fill=\dw, draw=gray] (3.25,\dx) rectangle (3.50,\dy); \node at (3.375,\dz) {\tiny \textcolor{white}{67}}; 
\filldraw[fill=\dw, draw=gray] (3.50,\dx) rectangle (3.75,\dy); \node at (3.625,\dz) {\tiny \textcolor{white}{68}}; 
\filldraw[fill=\aw, draw=gray] (3.75,\ax) rectangle (4.00,\ay); \node at (3.875,\az) {\tiny \textcolor{white}{69}}; 
\filldraw[fill=\dw, draw=gray] (4.00,\dx) rectangle (4.25,\dy); \node at (4.125,\dz) {\tiny \textcolor{white}{70}}; 
\filldraw[fill=\cw, draw=gray] (4.25,\cx) rectangle (4.50,\cy); \node at (4.375,\cz) {\tiny \textcolor{white}{71}}; 
\filldraw[fill=\bw, draw=gray] (4.50,\bx) rectangle (4.75,\by); \node at (4.625,\bz) {\tiny \textcolor{white}{72}}; 
\filldraw[fill=\cw, draw=gray] (4.75,\cx) rectangle (5.00,\cy); \node at (4.875,\cz) {\tiny \textcolor{white}{73}}; 
\filldraw[fill=\ew, draw=gray] (5.00,\ex) rectangle (5.25,\ey); \node at (5.125,\ez) {\tiny \textcolor{white}{74}}; 
\filldraw[fill=\cw, draw=gray] (5.25,\cx) rectangle (5.50,\cy); \node at (5.375,\cz) {\tiny \textcolor{white}{75}}; 
\filldraw[fill=\ew, draw=gray] (5.50,\ex) rectangle (5.75,\ey); \node at (5.625,\ez) {\tiny \textcolor{white}{76}}; 
\filldraw[fill=\aw, draw=gray] (5.75,\ax) rectangle (6.00,\ay); \node at (5.875,\az) {\tiny \textcolor{white}{77}}; 
\filldraw[fill=\ew, draw=gray] (6.00,\ex) rectangle (6.25,\ey); \node at (6.125,\ez) {\tiny \textcolor{white}{78}}; 
\filldraw[fill=\ew, draw=gray] (6.25,\ex) rectangle (6.50,\ey); \node at (6.375,\ez) {\tiny \textcolor{white}{79}}; 
\filldraw[fill=\bw, draw=gray] (6.50,\bx) rectangle (6.75,\by); \node at (6.625,\bz) {\tiny \textcolor{white}{80}}; 
\filldraw[fill=\bw, draw=gray] (6.75,\bx) rectangle (7.00,\by); \node at (6.875,\bz) {\tiny \textcolor{white}{81}}; 
\filldraw[fill=\ew, draw=gray] (7.00,\ex) rectangle (7.25,\ey); \node at (7.125,\ez) {\tiny \textcolor{white}{82}}; 
\filldraw[fill=\ew, draw=gray] (7.25,\ex) rectangle (7.50,\ey); \node at (7.375,\ez) {\tiny \textcolor{white}{83}}; 
\filldraw[fill=\aw, draw=gray] (7.50,\ax) rectangle (7.75,\ay); \node at (7.625,\az) {\tiny \textcolor{white}{84}}; 
\filldraw[fill=\ew, draw=gray] (7.75,\ex) rectangle (8.00,\ey); \node at (7.875,\ez) {\tiny \textcolor{white}{85}}; 
\filldraw[fill=\cw, draw=gray] (8.00,\cx) rectangle (8.25,\cy); \node at (8.125,\cz) {\tiny \textcolor{white}{86}}; 
\filldraw[fill=\ew, draw=gray] (8.25,\ex) rectangle (8.50,\ey); \node at (8.375,\ez) {\tiny \textcolor{white}{87}}; 
\filldraw[fill=\cw, draw=gray] (8.50,\cx) rectangle (8.75,\cy); \node at (8.625,\cz) {\tiny \textcolor{white}{88}}; 
\filldraw[fill=\bw, draw=gray] (8.75,\bx) rectangle (9.00,\by); \node at (8.875,\bz) {\tiny \textcolor{white}{89}}; 
\filldraw[fill=\cw, draw=gray] (9.00,\cx) rectangle (9.25,\cy); \node at (9.125,\cz) {\tiny \textcolor{white}{90}}; 
\filldraw[fill=\dw, draw=gray] (9.25,\dx) rectangle (9.50,\dy); \node at (9.375,\dz) {\tiny \textcolor{white}{91}}; 
\filldraw[fill=\aw, draw=gray] (9.50,\ax) rectangle (9.75,\ay); \node at (9.625,\az) {\tiny \textcolor{white}{92}}; 
\filldraw[fill=\dw, draw=gray] (9.75,\dx) rectangle (10.00,\dy); \node at (9.875,\dz) {\tiny \textcolor{white}{93}}; 
\filldraw[fill=\dw, draw=gray] (10.00,\dx) rectangle (10.25,\dy); \node at (10.125,\dz) {\tiny \textcolor{white}{94}}; 
\filldraw[fill=\ew, draw=gray] (10.25,\ex) rectangle (10.50,\ey); \node at (10.375,\ez) {\tiny \textcolor{white}{95}}; 
\filldraw[fill=\ew, draw=gray] (10.50,\ex) rectangle (10.75,\ey); \node at (10.625,\ez) {\tiny \textcolor{white}{96}}; 
\filldraw[fill=\aw, draw=gray] (10.75,\ax) rectangle (11.00,\ay); \node at (10.875,\az) {\tiny \textcolor{white}{97}}; 
\filldraw[fill=\bw, draw=gray] (11.00,\bx) rectangle (11.25,\by); \node at (11.125,\bz) {\tiny \textcolor{white}{98}}; 
\filldraw[fill=\bw, draw=gray] (11.25,\bx) rectangle (11.50,\by); \node at (11.375,\bz) {\tiny \textcolor{white}{99}}; 
\filldraw[fill=\dw, draw=gray] (11.50,\dx) rectangle (11.75,\dy); \node at (11.625,\dz) {\tiny \textcolor{white}{1\hspace{-0.5pt}00}}; 
\filldraw[fill=\cw, draw=gray] (11.75,\cx) rectangle (12.00,\cy); \node at (11.875,\cz) {\tiny \textcolor{white}{1\hspace{-0.5pt}01}}; 
\filldraw[fill=\cw, draw=gray] (12.00,\cx) rectangle (12.25,\cy); \node at (12.125,\cz) {\tiny \textcolor{white}{1\hspace{-0.5pt}02}}; 
\filldraw[fill=\bw, draw=gray] (12.25,\bx) rectangle (12.50,\by); \node at (12.375,\bz) {\tiny \textcolor{white}{1\hspace{-0.5pt}03}}; 
\filldraw[fill=\aw, draw=gray] (12.50,\ax) rectangle (12.75,\ay); \node at (12.625,\az) {\tiny \textcolor{white}{1\hspace{-0.5pt}04}}; 
\filldraw[fill=\dw, draw=gray] (12.75,\dx) rectangle (13.00,\dy); \node at (12.875,\dz) {\tiny \textcolor{white}{1\hspace{-0.5pt}05}}; 
\filldraw[fill=\dw, draw=gray] (13.00,\dx) rectangle (13.25,\dy); \node at (13.125,\dz) {\tiny \textcolor{white}{1\hspace{-0.5pt}06}}; 
\filldraw[fill=\aw, draw=gray] (13.25,\ax) rectangle (13.50,\ay); \node at (13.375,\az) {\tiny \textcolor{white}{1\hspace{-0.5pt}07}}; 
\end{tikzpicture}\\[0.2cm]
\begin{tikzpicture}[scale=1.31]
\draw[step=0.25cm,color=white] (-0.29,-0.25) grid (13.25,1);
\draw[step=0.25cm,color=gray] (0,-0.25) grid (13.25,1);
\filldraw[fill=\bw, draw=gray] (0,\bx) rectangle (0.25,\by); \node at (0.125,\bz) {\tiny \textcolor{white}{1\hspace{-0.5pt}08}}; 
\filldraw[fill=\bw, draw=gray] (0.25,\bx) rectangle (0.50,\by); \node at (0.375,\bz) {\tiny \textcolor{white}{1\hspace{-0.5pt}09}}; 
\filldraw[fill=\aw, draw=gray] (0.5,\ax) rectangle (0.75,\ay); \node at (0.625,\az) {\tiny \textcolor{white}{1\hspace{-0.5pt}10}}; 
\filldraw[fill=\cw, draw=gray] (0.75,\cx) rectangle (1.00,\cy); \node at (0.875,\cz) {\tiny \textcolor{white}{1\hspace{-0.5pt}11}}; 
\filldraw[fill=\aw, draw=gray] (1.00,\ax) rectangle (1.25,\ay); \node at (1.125,\az) {\tiny \textcolor{white}{1\hspace{-0.5pt}12}}; 
\filldraw[fill=\bw, draw=gray] (1.25,\bx) rectangle (1.50,\by); \node at (1.375,\bz) {\tiny \textcolor{white}{1\hspace{-0.5pt}13}}; 
\filldraw[fill=\cw, draw=gray] (1.50,\cx) rectangle (1.75,\cy); \node at (1.625,\cz) {\tiny \textcolor{white}{1\hspace{-0.5pt}14}}; 
\filldraw[fill=\cw, draw=gray] (1.75,\cx) rectangle (2.00,\cy); \node at (1.875,\cz) {\tiny \textcolor{white}{1\hspace{-0.5pt}15}}; 
\filldraw[fill=\ew, draw=gray] (2.00,\ex) rectangle (2.25,\ey); \node at (2.125,\ez) {\tiny \textcolor{white}{1\hspace{-0.5pt}16}}; 
\filldraw[fill=\bw, draw=gray] (2.25,\bx) rectangle (2.50,\by); \node at (2.375,\bz) {\tiny \textcolor{white}{1\hspace{-0.5pt}17}}; 
\filldraw[fill=\dw, draw=gray] (2.50,\dx) rectangle (2.75,\dy); \node at (2.625,\dz) {\tiny \textcolor{white}{1\hspace{-0.5pt}18}}; 
\filldraw[fill=\dw, draw=gray] (2.75,\dx) rectangle (3.00,\dy); \node at (2.875,\dz) {\tiny \textcolor{white}{1\hspace{-0.5pt}19}}; 
\filldraw[fill=\dw, draw=gray] (3.00,\dx) rectangle (3.25,\dy); \node at (3.125,\dz) {\tiny \textcolor{white}{1\hspace{-0.5pt}20}}; 
\filldraw[fill=\dw, draw=gray] (3.25,\dx) rectangle (3.50,\dy); \node at (3.375,\dz) {\tiny \textcolor{white}{1\hspace{-0.5pt}21}}; 
\filldraw[fill=\aw, draw=gray] (3.50,\ax) rectangle (3.75,\ay); \node at (3.625,\az) {\tiny \textcolor{white}{1\hspace{-0.5pt}22}}; 
\filldraw[fill=\bw, draw=gray] (3.75,\bx) rectangle (4.00,\by); \node at (3.875,\bz) {\tiny \textcolor{white}{1\hspace{-0.5pt}23}}; 
\filldraw[fill=\cw, draw=gray] (4.00,\cx) rectangle (4.25,\cy); \node at (4.125,\cz) {\tiny \textcolor{white}{1\hspace{-0.5pt}24}}; 
\filldraw[fill=\ew, draw=gray] (4.25,\ex) rectangle (4.50,\ey); \node at (4.375,\ez) {\tiny \textcolor{white}{1\hspace{-0.5pt}25}}; 
\filldraw[fill=\cw, draw=gray] (4.50,\cx) rectangle (4.75,\cy); \node at (4.625,\cz) {\tiny \textcolor{white}{1\hspace{-0.5pt}26}}; 
\filldraw[fill=\aw, draw=gray] (4.75,\ax) rectangle (5.00,\ay); \node at (4.875,\az) {\tiny \textcolor{white}{1\hspace{-0.5pt}27}}; 
\filldraw[fill=\cw, draw=gray] (5.00,\cx) rectangle (5.25,\cy); \node at (5.125,\cz) {\tiny \textcolor{white}{1\hspace{-0.5pt}28}}; 
\filldraw[fill=\ew, draw=gray] (5.25,\ex) rectangle (5.50,\ey); \node at (5.375,\ez) {\tiny \textcolor{white}{1\hspace{-0.5pt}29}}; 
\filldraw[fill=\aw, draw=gray] (5.50,\ax) rectangle (5.75,\ay); \node at (5.625,\az) {\tiny \textcolor{white}{1\hspace{-0.5pt}30}}; 
\filldraw[fill=\bw, draw=gray] (5.75,\bx) rectangle (6.00,\by); \node at (5.875,\bz) {\tiny \textcolor{white}{1\hspace{-0.5pt}31}}; 
\filldraw[fill=\aw, draw=gray] (6.00,\ax) rectangle (6.25,\ay); \node at (6.125,\az) {\tiny \textcolor{white}{1\hspace{-0.5pt}32}}; 
\filldraw[fill=\ew, draw=gray] (6.25,\ex) rectangle (6.50,\ey); \node at (6.375,\ez) {\tiny \textcolor{white}{1\hspace{-0.5pt}33}}; 
\filldraw[fill=\ew, draw=gray] (6.50,\ex) rectangle (6.75,\ey); \node at (6.625,\ez) {\tiny \textcolor{white}{1\hspace{-0.5pt}34}}; 
\filldraw[fill=\ew, draw=gray] (6.75,\ex) rectangle (7.00,\ey); \node at (6.875,\ez) {\tiny \textcolor{white}{1\hspace{-0.5pt}35}}; 
\filldraw[fill=\ew, draw=gray] (7.00,\ex) rectangle (7.25,\ey); \node at (7.125,\ez) {\tiny \textcolor{white}{1\hspace{-0.5pt}36}}; 
\filldraw[fill=\ew, draw=gray] (7.25,\ex) rectangle (7.50,\ey); \node at (7.375,\ez) {\tiny \textcolor{white}{1\hspace{-0.5pt}37}}; 
\filldraw[fill=\ew, draw=gray] (7.50,\ex) rectangle (7.75,\ey); \node at (7.625,\ez) {\tiny \textcolor{white}{1\hspace{-0.5pt}38}}; 
\filldraw[fill=\dw, draw=gray] (7.75,\dx) rectangle (8.00,\dy); \node at (7.875,\dz) {\tiny \textcolor{white}{1\hspace{-0.5pt}39}}; 
\filldraw[fill=\aw, draw=gray] (8.00,\ax) rectangle (8.25,\ay); \node at (8.125,\az) {\tiny \textcolor{white}{1\hspace{-0.5pt}40}}; 
\filldraw[fill=\cw, draw=gray] (8.25,\cx) rectangle (8.50,\cy); \node at (8.375,\cz) {\tiny \textcolor{white}{1\hspace{-0.5pt}41}}; 
\filldraw[fill=\aw, draw=gray] (8.50,\ax) rectangle (8.75,\ay); \node at (8.625,\az) {\tiny \textcolor{white}{1\hspace{-0.5pt}42}}; 
\filldraw[fill=\cw, draw=gray] (8.75,\cx) rectangle (9.00,\cy); \node at (8.875,\cz) {\tiny \textcolor{white}{1\hspace{-0.5pt}43}}; 
\filldraw[fill=\dw, draw=gray] (9.00,\dx) rectangle (9.25,\dy); \node at (9.125,\dz) {\tiny \textcolor{white}{1\hspace{-0.5pt}44}}; 
\filldraw[fill=\aw, draw=gray] (9.25,\ax) rectangle (9.50,\ay); \node at (9.375,\az) {\tiny \textcolor{white}{1\hspace{-0.5pt}45}}; 
\filldraw[fill=\ew, draw=gray] (9.50,\ex) rectangle (9.75,\ey); \node at (9.625,\ez) {\tiny \textcolor{white}{1\hspace{-0.5pt}46}}; 
\filldraw[fill=\aw, draw=gray] (9.75,\ax) rectangle (10.00,\ay); \node at (9.875,\az) {\tiny \textcolor{white}{1\hspace{-0.5pt}47}}; 
\filldraw[fill=\bw, draw=gray] (10.00,\bx) rectangle (10.25,\by); \node at (10.125,\bz) {\tiny \textcolor{white}{1\hspace{-0.5pt}48}}; 
\filldraw[fill=\bw, draw=gray] (10.25,\bx) rectangle (10.50,\by); \node at (10.375,\bz) {\tiny \textcolor{white}{1\hspace{-0.5pt}49}}; 
\filldraw[fill=\aw, draw=gray] (10.50,\ax) rectangle (10.75,\ay); \node at (10.625,\az) {\tiny \textcolor{white}{1\hspace{-0.5pt}50}}; 
\filldraw[fill=\dw, draw=gray] (10.75,\dx) rectangle (11.00,\dy); \node at (10.875,\dz) {\tiny \textcolor{white}{1\hspace{-0.5pt}51}}; 
\filldraw[fill=\aw, draw=gray] (11.00,\ax) rectangle (11.25,\ay); \node at (11.125,\az) {\tiny \textcolor{white}{1\hspace{-0.5pt}52}}; 
\filldraw[fill=\cw, draw=gray] (11.25,\cx) rectangle (11.50,\cy); \node at (11.375,\cz) {\tiny \textcolor{white}{1\hspace{-0.5pt}53}}; 
\filldraw[fill=\cw, draw=gray] (11.50,\cx) rectangle (11.75,\cy); \node at (11.625,\cz) {\tiny \textcolor{white}{1\hspace{-0.5pt}54}}; 
\filldraw[fill=\cw, draw=gray] (11.75,\cx) rectangle (12.00,\cy); \node at (11.875,\cz) {\tiny \textcolor{white}{1\hspace{-0.5pt}55}}; 
\filldraw[fill=\cw, draw=gray] (12.00,\cx) rectangle (12.25,\cy); \node at (12.125,\cz) {\tiny \textcolor{white}{1\hspace{-0.5pt}56}}; 
\filldraw[fill=\aw, draw=gray] (12.25,\ax) rectangle (12.50,\ay); \node at (12.375,\az) {\tiny \textcolor{white}{1\hspace{-0.5pt}57}}; 
\filldraw[fill=\bw, draw=gray] (12.50,\bx) rectangle (12.75,\by); \node at (12.625,\bz) {\tiny \textcolor{white}{1\hspace{-0.5pt}58}}; 
\filldraw[fill=\bw, draw=gray] (12.75,\bx) rectangle (13.00,\by); \node at (12.875,\bz) {\tiny \textcolor{white}{1\hspace{-0.5pt}59}}; 
\filldraw[fill=\aw, draw=gray] (13.00,\ax) rectangle (13.25,\ay); \node at (13.125,\az) {\tiny \textcolor{white}{1\hspace{-0.5pt}60}}; 
\end{tikzpicture}
\caption{Some extreme (and palindromic) certificates of the known Schur numbers and a palindromic certificate $S(5,160)$.}
\label{fig:extreme}
\end{figure*}




\section{Encoding}

To solve Schur Number Five, we first encode the existence of \emph{certificates} as propositional formulas
and then exploit the strength of a parallel SAT solver to efficiently determine whether these formulas are satisfiable.
A certificate $S(k,n)$ is a $k$-coloring of the numbers $1$ to $n$ with no monochromatic
solution of $a + b = c$ for $1 \leq a,b,c\leq n$. 
A certificate $S(k,n)$ provides a lower bound for the corresponding Schur problem: $S(k) \geq n$.
%
%
The size of a certificate $S(k,n)$ is $n$.
An {\em extreme certificate} is a certificate of maximum size.
Figure~\ref{fig:extreme} shows some extreme certificates for the known Schur numbers as 
well as a palindromic certificate $S(5,160)$ --- which is also an extreme certificate following the presented upper bound result. 
There is one extreme certificate modulo symmetry (i.e., modulo permuting the colors) with $k\in\{1,2\}$. These certificates are palindromes and thus modular.
There exist three extreme certificates $S(3,13)$ modulo symmetry. All of them
are modular and palindromes. They differ only regarding the color of number $7$, which can have any color. 
There are $273$ extreme certificates $S(4,44)$ modulo symmetry, of which $24$ are modular and palindromes~\cite{Palin}.

To establish that $S(k) = n$, we need to show that there exists a certificate $S(k,n)$ but no certificate $S(k,n+1)$.
We thus define a family of propositional formulas $F^k_n$, each of which encodes the existence of a certificate $S(k, n)$.
A satisfying assignment of $F^{5}_{160}$ can be computed in less than a minute by enforcing that the initial numbers cannot have the last color~\cite{Palin}.
The main challenge addressed in this paper is proving unsatisfiability of $F^{5}_{161}$ to show that $S(5) < 161$.
This requires many CPU years of computation even with optimized heuristics. 


For the formula $F^k_n$, we use Boolean variables
$\plit{j}{i}$ with $1 \leq i \leq k$ and $1 \leq j \leq n$. 
Intuitively, a variable $\plit{j}{i}$ is true if and only if number $j$ has color $i$ in the certificate.
The formula has three kinds of clauses: \emph{positive}, \emph{negative}, and \emph{optional}. 
The positive clauses encode that every number $j$ must have at least one color. They are of the form $(\plit{j}{1} \lor \dots \lor \plit{j}{k})$ for $1 \leq j \leq n$.
The negative clauses encode that for every solution of the equation $a + b = c$, the numbers $a$, $b$, and $c$ cannot have the same color $i$. They are of the form $(\nlit{a}{i} \lor \nlit{b}{i} \lor \nlit{c}{i})$
with $a+b=c$ and $1\leq a,b,c \leq n$.
Finally, the optional clauses encode that every number has at most one color. They are of the form $(\nlit{j}{h} \lor \nlit{j}{i})$ for $1 \leq h < i \leq k$. 
A commonly used SAT preprocessing technique, called
blocked clause elimination~\cite{BCE}, would remove the optional clauses. However, the optional clauses 
are required when counting or enumerating certificates.


\begin{example}
Formula $F^2_4$ consists of the following clauses:
\begin{eqnarray*}
\!\!\!\!\!\!&\!\!\!\!\!\!&(\plita{1}{1} \lor \plitb{1}{2}) \land (\plita{2}{1} \lor \plitb{2}{2}) \land (\plita{3}{1} \lor \plitb{3}{2}) \land (\plita{4}{1} \lor \plitb{4}{2})\,\land\\
\!\!\!\!\!\!&\!\!\!\!\!\!&(\nlita{1}{1} \lor \nlita{2}{1}) \land (\nlita{1}{1} \lor \nlita{2}{1} \lor \nlita{3}{1}) \land (\nlita{1}{1} \lor \nlita{3}{1} \lor \nlita{4}{1}) \land (\nlita{2}{1} \lor \nlita{4}{1})\,\land\\
\!\!\!\!\!\!&\!\!\!\!\!\!&(\nlitb{1}{2} \lor \nlitb{2}{2}) \land (\nlitb{1}{2} \lor \nlitb{2}{2} \lor \nlitb{3}{2}) \land (\nlitb{1}{2} \lor \nlitb{3}{2} \lor \nlitb{4}{2}) \land (\nlitb{2}{2} \lor \nlitb{4}{2})\,\land\\
\!\!\!\!\!\!&\!\!\!\!\!\!&(\nlita{1}{1} \lor \nlitb{1}{2}) \land (\nlita{2}{1} \lor \nlitb{2}{2}) \land (\nlita{3}{1} \lor \nlitb{3}{2}) \land (\nlita{4}{1} \lor \nlitb{4}{2})
\end{eqnarray*}

\noindent
The first line shows the positive clauses, the second and third line the negative clauses, and the last line the optional clauses. 
Notice that $(\nlita{1}{1} \lor \nlita{2}{1} \lor \nlita{3}{1})$ and $(\nlitb{1}{2} \lor \nlitb{2}{2} \lor \nlitb{3}{2})$ 
are subsumed by $(\nlita{1}{1} \lor \nlita{2}{1})$ and $(\nlitb{1}{2} \lor \nlitb{2}{2})$, respectively.

\end{example}



\subsection{Symmetry Breaking}


A certificate symmetry $\sigma$ for a Schur number problem is a mapping from
any certificate onto another certificate of that problem. 
Schur number problems have the certificate symmetry $\sigma_\mathrm{col}$ that permutes the colors.
%
Due to $\sigma_\mathrm{col}$, SAT solvers would explore all $5! = 120$ color permutations when solving formulas $F^5_n$. 
In the following, we describe how to fully and compactly break this symmetry by enforcing a lexicographical ordering on the colors~\cite{Crawford}.

Breaking the certificate symmetry $\sigma_{\mathrm{col}}$ for the first two colors is easy: 
We just assign the first color to number $1$ and the second color to number $2$.
Adding the unit clauses $(\plit{1}{1})$ and 
$(\plit{2}{2})$ to the formula will enforce this. 
Note that the two numbers must be colored differently because of the equation $1+1=2$.

Breaking $\sigma_{\mathrm{col}}$ for the third color is more involved.
At least one of the numbers $3$, $4$, and $5$ can have neither the first nor the second color due to $S(2) = 4$. 
We break the symmetry of the third color as follows: If number $4$ has neither the first nor the second color, we color it with the third color. Otherwise, if number $3$
has neither the first nor the second color, we color number $3$ with the third color. Otherwise, we color number 5 with the third color. 
We picked number $4$ as starting point as it is more constrained due to the equation $2+2=4$ and the clause $(\plit{2}{2})$. It therefore allows a more compact
symmetry-breaking predicate, which consists of the clauses
$(\nlit{3}{5})$, $(\nlit{4}{4})$, $(\nlit{4}{5})$, $(\plit{4}{3} \lor \nlit{3}{4})$, $(\plit{3}{4} \lor \nlit{5}{5})$, and $(\plit{3}{3} \lor \plit{4}{3} \lor \nlit{5}{4})$.
Finally, to distinguish between the fourth and the fifth color, we assign the fourth color to the first number that
does not have the first, second, or third color. We encode this with clauses of the form $(\plit{1}{4} \lor \dots \lor \plit{i}{4} \lor \nlit{i+1}{5})$.
We require these clauses only for $i \leq S(3) = 13$.

Generating the original formulas $F^k_n$ can be easily achieved with a dozen lines of code. In contrast, the addition of compact symmetry-breaking
predicates is more complicated and may therefore result in errors. Let $R^k_n$ be the formula obtained from $F^k_n$ by adding symmetry-breaking predicates.
To ensure correctness of the symmetry breaking, we constructed
a proof, called the {\em re-encoding proof}, that the satisfiability of $F^{5}_{161}$ implies the satisfiability of $R^{5}_{161}$.

\section{Decision Heuristics}

We used the \emph{cube-and-conquer} method~\cite{HKWB11} for SAT solving as
it is arguably the most effective method for solving very hard combinatorial problems. 
This method was also used for solving the Erd\H{o}s discrepancy problem~\cite{Konev:2015} and the Pythagorean
triples problem~\cite{ptn}.

Cube-and-conquer is a hybrid parallel SAT solving paradigm that combines \emph{look-ahead} techniques~\cite{HvM09HBSAT}
with \emph{conflict-driven clause learning} (CDCL)~\cite{MSLM09HBSAT}: Look-ahead techniques are used for splitting a given problem into many (millions or even billions of) subproblems which are then solved with CDCL solvers. 
Since the subproblems are
independent, they can be easily solved in parallel without requiring communication.

The aim of look-ahead techniques is to find variable assignments that simplify a formula
as much as possible.
This is achieved with so-called \emph{look-aheads}: A look-ahead on a literal $l$ with respect to a formula $F$
first assigns $l$ to true and then simplifies $F$ to obtain a formula $F'$. 
After this, it determines a heuristic value by computing
the ``difference'' between $F$ and $F'$ (details are given below). A variable $v$ is considered 
useful for splitting a formula $F$ if the look-aheads on both $v$ and $\overline v$ have
a high heuristic value. Typically, look-ahead techniques select the variable $v$ for which the product of the heuristic values of $v$ and $\overline v$ is the largest. 

The effectiveness of look-ahead heuristics depends on measuring the difference between
the formula $F$ and the simplified formula $F'$. A reasonably effective measure, which is also easy to compute, 
is the difference in the number of variables: $\left|\var(F)\right|$$-$$\left|\var(F')\right|$. This measure
is used in the cube-and-conquer solver {\sc Treengeling}~\cite{biere2013}, which solved most benchmarks of the
SAT Competition 2016~\cite{SC16}. An alternative, more costly measure, considers the 
clauses that have been reduced, but not satisfied, during the simplification, i.e., the clauses in $F' \setminus F$.
These clauses are typically assigned a weight, with shorter clauses getting a larger
weight. During our initial experiments for solving Schur Number Five, we observed that 
the clause-based heuristics is much more effective than the variable-based one.
However, our initial experiments---based on splitting the problem into millions of subproblems and 
solving randomly selected subproblems---indicated that finding the solution of Schur Number Five would require
many decades of CPU time. 

The key to reducing the computational effort of solving Schur Number Five is a new measurement method. We first discuss the main
weakness of the weighted-sum heuristics before we describe our new method.
Recall that the Schur number encoding
uses $\mathcal{O}(n)$ positive clauses of length $k$ and $\mathcal{O}(kn^2)$ negative
clauses of length $3$. Thus, no matter on what literal we look ahead, most clauses in $F' \setminus F$ originate from negative clauses.
Moreover, a clause in $F' \setminus F$ that originates from a negative clause has length $2$, while a
clause in $F' \setminus F$ that originates from a positive clause can be larger. Commonly used heuristics
favor shorter clauses and thus favor clauses that originate from negative clauses. Because of this, the heuristic
value of look-aheads is dominated by reduced negative clauses. However, it appears that favoring
reduced positive clauses is more effective.

\begin{example}
Recall $F^2_4$, but now without redundant clauses: 
\begin{eqnarray*}
\!\!\!\!\!\!&\!\!\!\!\!\!&(\plita{1}{1} \lor \plitb{1}{2}) \land (\plita{2}{1} \lor \plitb{2}{2}) \land (\plita{3}{1} \lor \plitb{3}{2}) \land (\plita{4}{1} \lor \plitb{4}{2})\,\land\\
\!\!\!\!\!\!&\!\!\!\!\!\!&(\nlita{1}{1} \lor \nlita{2}{1}) \land (\nlita{1}{1} \lor \nlita{3}{1} \lor \nlita{4}{1}) \land (\nlita{2}{1} \lor \nlita{4}{1})\,\land\\
\!\!\!\!\!\!&\!\!\!\!\!\!&(\nlitb{1}{2} \lor \nlitb{2}{2}) \land (\nlitb{1}{2} \lor \nlitb{3}{2} \lor \nlitb{4}{2}) \land (\nlitb{2}{2} \lor \nlitb{4}{2})
\end{eqnarray*}
Let us look ahead on literal $\nlita{3}{1}$: Assigning variable $\plita{3}{1}$ to false satisfies $(\nlita{1}{1} \lor \nlita{3}{1} \lor \nlita{4}{1})$
and reduces $(\plita{3}{1} \lor \plitb{3}{2})$ to $(\plitb{3}{2})$, thereby forcing the variable $\plitb{3}{2}$ to true. This in turn
reduces the negative clause $(\nlitb{1}{2} \lor \nlitb{3}{2} \lor \nlitb{4}{2})$ to $(\nlitb{1}{2} \lor \nlitb{4}{2})$. The only clause that is reduced, but not 
satisfied, is $(\nlitb{1}{2} \lor \nlitb{4}{2})$. 
Hence, looking ahead on $\nlita{3}{1}$ yields $F' \setminus F = (\nlitb{1}{2} \lor \nlitb{4}{2})$.
\end{example}

\noindent
We now present our generalization of an effective heuristic for uniform random 3-SAT instances~\cite{Li:1999} to arbitrary CNFs. 
Given a literal $l$ and a formula $F$, let $occ(F,l)$ denote the number of occurrences of $l$ in $F$. The weight of a clause 
$C \in F$, denoted by $w(F,C)$, is computed as follows:
\[
w(F,C) = \frac{\sum_{l \in C}occ(F,\overline l)}{2^{|C|}\cdot|C|}
\]

\noindent
The $|C|$ in the denominator reduces the sum to the average and $2^{|C|}$ ensures a larger weight for shorter clauses. 
We noticed that the sum works much better than the product for arbitrary CNFs, in contrast to random 3-SAT formulas~\cite{kcnfs}.
The heuristic value of a variable $v$ w.r.t. a formula $F$, denoted by $H(F,v)$, is computed as follows (with $F'$ and $F''$ referring to the
formulas obtained by look-aheads on $F$ with the literals $v$ and $\overline v$, respectively):
\[
H(F,v) = \big(\!\!\!\sum_{C\in F'\setminus F} \!\!\!\!\!w(F,C)\big) \cdot \big(\!\!\!\sum_{C\in F''\setminus F} \!\!\!\!\!w(F,C)\big)
\]
\noindent
In each node of the search tree, the variable with the highest heuristic value 
is selected as splitting variable. 

\section{Partitioning}

A crucial part of solving Schur Number Five is the partitioning of the propositional formulas $R^5_{160}$ and
$R^5_{161}$ into millions of easy subproblems. We use the former formula to compute all extreme certificates $S(5,160)$
and the latter formula for the upper bound result. We constructed a single partition for both formulas.
A partition is a set of cubes (or equivalently, a set of variable assignments). The disjunction of cubes in a partition must
be a tautology in order to ensure that the cubes cover the entire search space. By applying a cube to a formula, 
one obtains a subproblem of that formula. 
Each of the subproblems arising from a partition can be solved in parallel, thereby allowing massively
parallel computation. Moreover, these subproblems are partitioned again to solve them more efficiently (on a single core). 

The top-level partition is constructed as follows: We use the look-ahead decision heuristic described above to build a binary search tree 
over the space of possible variable assignments. In this tree, every non-leaf node is assigned a splitting variable. The left outgoing edge of a node assigns
its variable to true while the right one assigns it to false. 
Each node in the tree represents the variable assignment corresponding to all assignments on the path from the root node. 
In case the formula in a node (i.e., the formula obtained from the original formula by applying the assignment represented by that node) becomes ``easy'', we stop splitting.
The partition consists of all assignments that are represented by the leaf nodes of the tree.
We require a measure that captures the hardness of a formula in each node.
A rough measure suffices here, since we will fine-tune this partition later. 
We observed that the number of binary clauses in a formula is a reasonable measure for the hardness of Schur number subproblems. 
The more binary clauses, the more constrained the
subproblem (and thus easier to solve). For example, stopping with the splitting as soon as a
formula in a node has more than $3700$ binary clauses results in about $9$ millions of mostly easy subproblems of 
$R^5_{160}$ and $R^5_{161}$. 

\subsection{The Hidden Strength of Cube-and-Conquer}

Cube-and-conquer is not only useful for partitioning a hard problem into many subproblems that can be solved in parallel, 
but also to boost performance of solving a problem on a single core. 
Let $N$ be the number of cubes in a partition.
A low value of $N$ indicates that the problem is split into a low number of subproblems, meaning that it is mainly solved with CDCL ($N = 1$ means pure CDCL) while a larger value indicates a more extensive splitting based on look-aheads.

If we experiment with different values for $N$ when trying to solve a problem on a single core, 
we can observe an interesting pattern:
For low values of $N$, an increase of $N$ leads to an increase of the total runtime---apparently 
some subproblems are about as hard as the original one.
If we increase $N$ further, the total runtime starts to decrease and at some point it can even become
significantly smaller compared to solving the problem with CDCL alone (again running both on a single core). 
Yet when
$N$ becomes really large, the runtime increases again. At this point, splitting starts to dominate 
the total costs. Figure~\ref{fig:hidden} shows this pattern on a subproblem of $R^5_{161}$,
where the optimal value for $N$ is around $10\,000$.

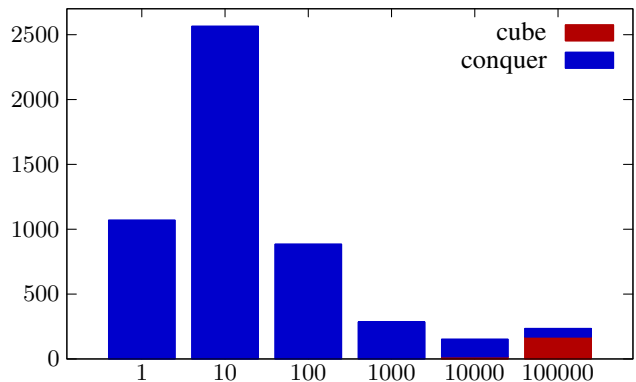
\begin{figure}[h]
\centering
\begin{tikzpicture}[xscale=0.7,yscale=0.6,gnuplot]
\path (1.000,0.000) rectangle (12.500,8.750);
\gpcolor{color=gp lt color border}
\gpsetlinetype{gp lt border}
\gpsetdashtype{gp dt solid}
\gpsetlinewidth{1.00}
\draw[gp path] (1.196,0.616)--(1.376,0.616);
\draw[gp path] (11.947,0.616)--(11.767,0.616);
\node[gp node right] at (1.012,0.616) {\footnotesize $0$\!\!\!};
\draw[gp path] (1.196,2.054)--(1.376,2.054);
\draw[gp path] (11.947,2.054)--(11.767,2.054);
\node[gp node right] at (1.012,2.054) {\footnotesize $500$\!\!\!};
\draw[gp path] (1.196,3.492)--(1.376,3.492);
\draw[gp path] (11.947,3.492)--(11.767,3.492);
\node[gp node right] at (1.012,3.492) {\footnotesize $1000$\!\!\!};
\draw[gp path] (1.196,4.930)--(1.376,4.930);
\draw[gp path] (11.947,4.930)--(11.767,4.930);
\node[gp node right] at (1.012,4.930) {\footnotesize $1500$\!\!\!};
\draw[gp path] (1.196,6.368)--(1.376,6.368);
\draw[gp path] (11.947,6.368)--(11.767,6.368);
\node[gp node right] at (1.012,6.368) {\footnotesize $2000$\!\!\!};
\draw[gp path] (1.196,7.806)--(1.376,7.806);
\draw[gp path] (11.947,7.806)--(11.767,7.806);
\node[gp node right] at (1.012,7.806) {\footnotesize $2500$\!\!\!};
\draw[gp path] (2.619,0.616)--(2.619,0.796);
\draw[gp path] (2.619,8.381)--(2.619,8.201);
\node[gp node center] at (2.619,0.308) {\footnotesize$1$};
\draw[gp path] (4.200,0.616)--(4.200,0.796);
\draw[gp path] (4.200,8.381)--(4.200,8.201);
\node[gp node center] at (4.200,0.308) {\footnotesize$10$};
\draw[gp path] (5.781,0.616)--(5.781,0.796);
\draw[gp path] (5.781,8.381)--(5.781,8.201);
\node[gp node center] at (5.781,0.308) {\footnotesize$100$};
\draw[gp path] (7.362,0.616)--(7.362,0.796);
\draw[gp path] (7.362,8.381)--(7.362,8.201);
\node[gp node center] at (7.362,0.308) {\footnotesize $1000$};
\draw[gp path] (8.943,0.616)--(8.943,0.796);
\draw[gp path] (8.943,8.381)--(8.943,8.201);
\node[gp node center] at (8.943,0.308) {\footnotesize $10000$};
\draw[gp path] (10.524,0.616)--(10.524,0.796);
\draw[gp path] (10.524,8.381)--(10.524,8.201);
\node[gp node center] at (10.524,0.308) {\footnotesize $100000$};
\draw[gp path] (1.196,8.381)--(1.196,0.616)--(11.947,0.616)--(11.947,8.381)--cycle;
\node[gp node right] at (10.479,7.847) {cube};
\gpfill{rgb color={\colorb}} (10.663,7.670)--(11.579,7.670)--(11.579,8.024)--(10.663,8.024)--cycle;
\gpcolor{rgb color={\colorb}}
\draw[gp path] (10.663,7.670)--(11.579,7.670)--(11.579,8.024)--(10.663,8.024)--cycle;
\gpfill{rgb color={\colorb}} (8.311,0.616)--(9.576,0.616)--(9.576,0.666)--(8.311,0.666)--cycle;
\draw[gp path] (8.311,0.616)--(8.311,0.665)--(9.575,0.665)--(9.575,0.616)--cycle;
\gpfill{rgb color={\colorb}} (9.892,0.616)--(11.157,0.616)--(11.157,1.106)--(9.892,1.106)--cycle;
\draw[gp path] (9.892,0.616)--(9.892,1.105)--(11.156,1.105)--(11.156,0.616)--cycle;
\gpcolor{color=gp lt color border}
\node[gp node right] at (10.479,7.239) {conquer};
\gpfill{rgb color={\colora}} (10.663,7.062)--(11.579,7.062)--(11.579,7.416)--(10.663,7.416)--cycle;
\gpcolor{rgb color={\colora}}
\draw[gp path] (10.663,7.062)--(11.579,7.062)--(11.579,7.416)--(10.663,7.416)--cycle;
\gpfill{rgb color={\colora}} (1.987,0.616)--(3.252,0.616)--(3.252,3.694)--(1.987,3.694)--cycle;
\draw[gp path] (1.987,0.616)--(1.987,3.693)--(3.251,3.693)--(3.251,0.616)--cycle;
\gpfill{rgb color={\colora}} (3.568,0.616)--(4.833,0.616)--(4.833,7.994)--(3.568,7.994)--cycle;
\draw[gp path] (3.568,0.616)--(3.568,7.993)--(4.832,7.993)--(4.832,0.616)--cycle;
\gpfill{rgb color={\colora}} (5.149,0.616)--(6.414,0.616)--(6.414,3.162)--(5.149,3.162)--cycle;
\draw[gp path] (5.149,0.616)--(5.149,3.161)--(6.413,3.161)--(6.413,0.616)--cycle;
\gpfill{rgb color={\colora}} (6.730,0.616)--(7.995,0.616)--(7.995,1.437)--(6.730,1.437)--cycle;
\draw[gp path] (6.730,0.616)--(6.730,1.436)--(7.994,1.436)--(7.994,0.616)--cycle;
\gpfill{rgb color={\colora}} (8.311,0.665)--(9.576,0.665)--(9.576,1.054)--(8.311,1.054)--cycle;
\draw[gp path] (8.311,0.665)--(8.311,1.053)--(9.575,1.053)--(9.575,0.665)--cycle;
\gpfill{rgb color={\colora}} (9.892,1.105)--(11.157,1.105)--(11.157,1.293)--(9.892,1.293)--cycle;
\draw[gp path] (9.892,1.105)--(9.892,1.292)--(11.156,1.292)--(11.156,1.105)--cycle;
\gpcolor{color=gp lt color border}
\draw[gp path] (1.196,8.381)--(1.196,0.616)--(11.947,0.616)--(11.947,8.381)--cycle;
\gpdefrectangularnode{gp plot 1}{\pgfpoint{1.196cm}{0.616cm}}{\pgfpoint{11.947cm}{8.381cm}}
\end{tikzpicture}
\caption{Comparison of the total runtime in seconds (\mbox{$y$-a}xis) for solving a subproblem of $R^5_{161}$ using different numbers of cubes ($x$-axis) on a single core (no parallelism).}
\label{fig:hidden}
\end{figure}

We developed a mechanism that approximates the optimal number to realize fast performance.
The mechanism stops splitting if the number of remaining variables in a node drops below the value of parameter $\delta$. 
The number of remaining variables is a useful measure as it strictly decreases, whereas
number of binary clauses can oscillate. Initial experiments showed that such oscillation can slowdown the solver on easy problems.
We initialize $\delta$ to $0$, meaning that we keep splitting until $\delta$ is increased. We only increase $\delta$ when look-ahead techniques
can refute a node, which naturally terminates splitting. In this case, $\delta$ is set to the number of remaining variables in that node. The
increase is motivated as follows: If look-ahead techniques can refute a node, then we expect CDCL to
refute that node---as well as similar nodes---more efficiently.

The value of $\delta$ is decreased in each node of the search tree to ensure that look-ahead techniques
refute nodes once in a while. We experimented with various 
methods to implement the decrement and observed that the size of the decrease should 
be related to the depth of a node in the search tree. The closer a node is to the root, the larger the decrement.
More specifically, we used the following update function (with $d$ referring to the depth of the node that performs the update and 
parameter $e$ referring to  {\em down exponent} and parameter $f$ referring to the {\em down factor}): 
\[
\delta := \delta \big(1 - f^{d^e}\big)
\]
If the value of $f$ is close to $0$, then $\delta$ climbs to a value at which look-ahead techniques 
will rarely refute a node. On the other hand, if the value of $f$ is close to $1$, then $\delta$ drops quickly 
to $0$, so that practically all leaf nodes will be refuted by look-ahead techniques. The value of $e$ determines
the influence of the depth. If $e$ is close to $0$, then the depth is ignored during the update, while if $e$ is
close to $1$ (or even larger), then the depth is dominant.  
During our experiments, various combinations of values of $e$ and $f$ resulted in strong performance. Examples 
of effective values are $e = 1.0$ and $f = 0.6$, $e = 0.5$ and $f = 0.1$, and $e = 0.3$ and $f = 0.02$.
For the final experiments we used $e = 0.3$ and $f = 0. 02$.

\subsection{Hardness Predictor and Partition Balancing}

Most modern microprocessors used in clusters, including the Intel Xeon chips we used, 
have many CPU cores yet relatively little memory---at least for our application, which runs many SAT solvers and theorem provers in parallel.
A major challenge was technical in nature: maximizing the CPU usage (with hyper-threading) without running
out of memory. 

Although most subproblems generated in the top-level partition could be solved within 
reasonable time (less than two minutes) some subproblems required hours of computation. 
Moreover, solving these hard subproblems required disproportionally more memory. Solving a few hard problems
on the same chip at the same time could kill all threads. We therefore fine-tuned the initial partition.

We used the mechanism to partition subproblems as a hardness predictor of  subproblems by using
a high down exponent and a  small down factor: $e=1.0$ and $f = 0.1$. Using a small down factor boosts the partitioning runtime, but
produces typically too few cubes. However, here we only care about the runtime.
It turned out that the runtime of partitioning with $e=1.0$ and $f = 0.1$ is larger than a second for hard subproblems and
significantly smaller for the easier ones. We extended the partition by splitting subproblems if this hardness predictor took over a second. Splitting was continued until
none of the subproblems was predicted to be hard. To limit the size of the partition,
we merged two cubes if they had the same parent node and the sum of their hardness predictor times was
less than $0.1$ second.

%

\begin{figure*}[h]
\begin{tikzpicture}[scale=0.55,gnuplot]
\path (1.000,0.000) rectangle (12.500,8.750);
\gpcolor{color=gp lt color border}
\gpsetlinetype{gp lt border}
\gpsetdashtype{gp dt solid}
\gpsetlinewidth{1.00}
\draw[gp path] (2.965,0.616)--(3.145,0.616);
\draw[gp path] (10.730,0.616)--(10.550,0.616);
\node[gp node right] at (2.781,0.616) {\scriptsize $0$};
\draw[gp path] (2.965,1.733)--(3.145,1.733);
\draw[gp path] (10.730,1.733)--(10.550,1.733);
\node[gp node right] at (2.781,1.733) {\scriptsize $0.2\!\cdot\!10^{6}\!\!\!$};
\draw[gp path] (2.965,2.851)--(3.145,2.851);
\draw[gp path] (10.730,2.851)--(10.550,2.851);
\node[gp node right] at (2.781,2.851) {\scriptsize $0.4\!\cdot\!10^{6}\!\!\!$};
\draw[gp path] (2.965,3.968)--(3.145,3.968);
\draw[gp path] (10.730,3.968)--(10.550,3.968);
\node[gp node right] at (2.781,3.968) {\scriptsize $0.6\!\cdot\!10^{6}\!\!\!$};
\draw[gp path] (2.965,5.085)--(3.145,5.085);
\draw[gp path] (10.730,5.085)--(10.550,5.085);
\node[gp node right] at (2.781,5.085) {\scriptsize $0.8\!\cdot\!10^{6}\!\!\!$};
\draw[gp path] (2.965,6.202)--(3.145,6.202);
\draw[gp path] (10.730,6.202)--(10.550,6.202);
\node[gp node right] at (2.781,6.202) {\scriptsize $1.0\!\cdot\!10^{6}\!\!\!$};
\draw[gp path] (2.965,7.320)--(3.145,7.320);
\draw[gp path] (10.730,7.320)--(10.550,7.320);
\node[gp node right] at (2.781,7.320) {\scriptsize $1.2\!\cdot\!10^{6}\!\!\!$};
\draw[gp path] (4.183,0.616)--(4.183,0.796);
\draw[gp path] (4.183,8.381)--(4.183,8.201);
\node[gp node center] at (4.183,0.308) {\scriptsize $20$};
\draw[gp path] (5.706,0.616)--(5.706,0.796);
\draw[gp path] (5.706,8.381)--(5.706,8.201);
\node[gp node center] at (5.706,0.308) {\scriptsize $30$};
\draw[gp path] (7.228,0.616)--(7.228,0.796);
\draw[gp path] (7.228,8.381)--(7.228,8.201);
\node[gp node center] at (7.228,0.308) {\scriptsize $40$};
\draw[gp path] (8.751,0.616)--(8.751,0.796);
\draw[gp path] (8.751,8.381)--(8.751,8.201);
\node[gp node center] at (8.751,0.308) {\scriptsize $50$};
\draw[gp path] (10.273,0.616)--(10.273,0.796);
\draw[gp path] (10.273,8.381)--(10.273,8.201);
\node[gp node center] at (10.273,0.308) {\scriptsize $60$};
\draw[gp path] (2.965,8.381)--(2.965,0.616)--(10.730,0.616)--(10.730,8.381)--cycle;
\gpfill{rgb color={0.580,0.000,0.827}} (3.064,0.616)--(3.172,0.616)--(3.172,0.617)--(3.064,0.617)--cycle;
\gpcolor{rgb color={\colora}}
\draw[gp path] (3.064,0.616)--(3.171,0.616)--cycle;
\gpfill{rgb color={\colora}} (3.216,0.616)--(3.324,0.616)--(3.324,0.617)--(3.216,0.617)--cycle;
\draw[gp path] (3.216,0.616)--(3.323,0.616)--cycle;
\gpfill{rgb color={\colora}} (3.368,0.616)--(3.476,0.616)--(3.476,0.619)--(3.368,0.619)--cycle;
\draw[gp path] (3.368,0.616)--(3.368,0.618)--(3.475,0.618)--(3.475,0.616)--cycle;
\gpfill{rgb color={\colora}} (3.521,0.616)--(3.628,0.616)--(3.628,0.630)--(3.521,0.630)--cycle;
\draw[gp path] (3.521,0.616)--(3.521,0.629)--(3.627,0.629)--(3.627,0.616)--cycle;
\gpfill{rgb color={\colora}} (3.673,0.616)--(3.781,0.616)--(3.781,0.668)--(3.673,0.668)--cycle;
\draw[gp path] (3.673,0.616)--(3.673,0.667)--(3.780,0.667)--(3.780,0.616)--cycle;
\gpfill{rgb color={\colora}} (3.825,0.616)--(3.933,0.616)--(3.933,0.776)--(3.825,0.776)--cycle;
\draw[gp path] (3.825,0.616)--(3.825,0.775)--(3.932,0.775)--(3.932,0.616)--cycle;
\gpfill{rgb color={\colora}} (3.977,0.616)--(4.085,0.616)--(4.085,1.030)--(3.977,1.030)--cycle;
\draw[gp path] (3.977,0.616)--(3.977,1.029)--(4.084,1.029)--(4.084,0.616)--cycle;
\gpfill{rgb color={\colora}} (4.130,0.616)--(4.237,0.616)--(4.237,1.522)--(4.130,1.522)--cycle;
\draw[gp path] (4.130,0.616)--(4.130,1.521)--(4.236,1.521)--(4.236,0.616)--cycle;
\gpfill{rgb color={\colora}} (4.282,0.616)--(4.390,0.616)--(4.390,2.323)--(4.282,2.323)--cycle;
\draw[gp path] (4.282,0.616)--(4.282,2.322)--(4.389,2.322)--(4.389,0.616)--cycle;
\gpfill{rgb color={\colora}} (4.434,0.616)--(4.542,0.616)--(4.542,3.433)--(4.434,3.433)--cycle;
\draw[gp path] (4.434,0.616)--(4.434,3.432)--(4.541,3.432)--(4.541,0.616)--cycle;
\gpfill{rgb color={\colora}} (4.587,0.616)--(4.694,0.616)--(4.694,4.753)--(4.587,4.753)--cycle;
\draw[gp path] (4.587,0.616)--(4.587,4.752)--(4.693,4.752)--(4.693,0.616)--cycle;
\gpfill{rgb color={\colora}} (4.739,0.616)--(4.846,0.616)--(4.846,6.012)--(4.739,6.012)--cycle;
\draw[gp path] (4.739,0.616)--(4.739,6.011)--(4.845,6.011)--(4.845,0.616)--cycle;
\gpfill{rgb color={\colora}} (4.891,0.616)--(4.999,0.616)--(4.999,6.927)--(4.891,6.927)--cycle;
\draw[gp path] (4.891,0.616)--(4.891,6.926)--(4.998,6.926)--(4.998,0.616)--cycle;
\gpfill{rgb color={\colora}} (5.043,0.616)--(5.151,0.616)--(5.151,7.282)--(5.043,7.282)--cycle;
\draw[gp path] (5.043,0.616)--(5.043,7.281)--(5.150,7.281)--(5.150,0.616)--cycle;
\gpfill{rgb color={\colora}} (5.196,0.616)--(5.303,0.616)--(5.303,6.983)--(5.196,6.983)--cycle;
\draw[gp path] (5.196,0.616)--(5.196,6.982)--(5.302,6.982)--(5.302,0.616)--cycle;
\gpfill{rgb color={\colora}} (5.348,0.616)--(5.455,0.616)--(5.455,6.172)--(5.348,6.172)--cycle;
\draw[gp path] (5.348,0.616)--(5.348,6.171)--(5.454,6.171)--(5.454,0.616)--cycle;
\gpfill{rgb color={\colora}} (5.500,0.616)--(5.608,0.616)--(5.608,5.076)--(5.500,5.076)--cycle;
\draw[gp path] (5.500,0.616)--(5.500,5.075)--(5.607,5.075)--(5.607,0.616)--cycle;
\gpfill{rgb color={\colora}} (5.652,0.616)--(5.760,0.616)--(5.760,3.952)--(5.652,3.952)--cycle;
\draw[gp path] (5.652,0.616)--(5.652,3.951)--(5.759,3.951)--(5.759,0.616)--cycle;
\gpfill{rgb color={\colora}} (5.805,0.616)--(5.912,0.616)--(5.912,2.995)--(5.805,2.995)--cycle;
\draw[gp path] (5.805,0.616)--(5.805,2.994)--(5.911,2.994)--(5.911,0.616)--cycle;
\gpfill{rgb color={\colora}} (5.957,0.616)--(6.064,0.616)--(6.064,2.273)--(5.957,2.273)--cycle;
\draw[gp path] (5.957,0.616)--(5.957,2.272)--(6.063,2.272)--(6.063,0.616)--cycle;
\gpfill{rgb color={\colora}} (6.109,0.616)--(6.217,0.616)--(6.217,1.787)--(6.109,1.787)--cycle;
\draw[gp path] (6.109,0.616)--(6.109,1.786)--(6.216,1.786)--(6.216,0.616)--cycle;
\gpfill{rgb color={\colora}} (6.261,0.616)--(6.369,0.616)--(6.369,1.488)--(6.261,1.488)--cycle;
\draw[gp path] (6.261,0.616)--(6.261,1.487)--(6.368,1.487)--(6.368,0.616)--cycle;
\gpfill{rgb color={\colora}} (6.414,0.616)--(6.521,0.616)--(6.521,1.299)--(6.414,1.299)--cycle;
\draw[gp path] (6.414,0.616)--(6.414,1.298)--(6.520,1.298)--(6.520,0.616)--cycle;
\gpfill{rgb color={\colora}} (6.566,0.616)--(6.673,0.616)--(6.673,1.169)--(6.566,1.169)--cycle;
\draw[gp path] (6.566,0.616)--(6.566,1.168)--(6.672,1.168)--(6.672,0.616)--cycle;
\gpfill{rgb color={\colora}} (6.718,0.616)--(6.826,0.616)--(6.826,1.072)--(6.718,1.072)--cycle;
\draw[gp path] (6.718,0.616)--(6.718,1.071)--(6.825,1.071)--(6.825,0.616)--cycle;
\gpfill{rgb color={\colora}} (6.870,0.616)--(6.978,0.616)--(6.978,0.989)--(6.870,0.989)--cycle;
\draw[gp path] (6.870,0.616)--(6.870,0.988)--(6.977,0.988)--(6.977,0.616)--cycle;
\gpfill{rgb color={\colora}} (7.023,0.616)--(7.130,0.616)--(7.130,0.910)--(7.023,0.910)--cycle;
\draw[gp path] (7.023,0.616)--(7.023,0.909)--(7.129,0.909)--(7.129,0.616)--cycle;
\gpfill{rgb color={\colora}} (7.175,0.616)--(7.282,0.616)--(7.282,0.843)--(7.175,0.843)--cycle;
\draw[gp path] (7.175,0.616)--(7.175,0.842)--(7.281,0.842)--(7.281,0.616)--cycle;
\gpfill{rgb color={\colora}} (7.327,0.616)--(7.435,0.616)--(7.435,0.789)--(7.327,0.789)--cycle;
\draw[gp path] (7.327,0.616)--(7.327,0.788)--(7.434,0.788)--(7.434,0.616)--cycle;
\gpfill{rgb color={\colora}} (7.479,0.616)--(7.587,0.616)--(7.587,0.746)--(7.479,0.746)--cycle;
\draw[gp path] (7.479,0.616)--(7.479,0.745)--(7.586,0.745)--(7.586,0.616)--cycle;
\gpfill{rgb color={\colora}} (7.632,0.616)--(7.739,0.616)--(7.739,0.715)--(7.632,0.715)--cycle;
\draw[gp path] (7.632,0.616)--(7.632,0.714)--(7.738,0.714)--(7.738,0.616)--cycle;
\gpfill{rgb color={\colora}} (7.784,0.616)--(7.891,0.616)--(7.891,0.695)--(7.784,0.695)--cycle;
\draw[gp path] (7.784,0.616)--(7.784,0.694)--(7.890,0.694)--(7.890,0.616)--cycle;
\gpfill{rgb color={\colora}} (7.936,0.616)--(8.044,0.616)--(8.044,0.680)--(7.936,0.680)--cycle;
\draw[gp path] (7.936,0.616)--(7.936,0.679)--(8.043,0.679)--(8.043,0.616)--cycle;
\gpfill{rgb color={\colora}} (8.088,0.616)--(8.196,0.616)--(8.196,0.669)--(8.088,0.669)--cycle;
\draw[gp path] (8.088,0.616)--(8.088,0.668)--(8.195,0.668)--(8.195,0.616)--cycle;
\gpfill{rgb color={\colora}} (8.241,0.616)--(8.348,0.616)--(8.348,0.660)--(8.241,0.660)--cycle;
\draw[gp path] (8.241,0.616)--(8.241,0.659)--(8.347,0.659)--(8.347,0.616)--cycle;
\gpfill{rgb color={\colora}} (8.393,0.616)--(8.500,0.616)--(8.500,0.651)--(8.393,0.651)--cycle;
\draw[gp path] (8.393,0.616)--(8.393,0.650)--(8.499,0.650)--(8.499,0.616)--cycle;
\gpfill{rgb color={\colora}} (8.545,0.616)--(8.653,0.616)--(8.653,0.645)--(8.545,0.645)--cycle;
\draw[gp path] (8.545,0.616)--(8.545,0.644)--(8.652,0.644)--(8.652,0.616)--cycle;
\gpfill{rgb color={\colora}} (8.697,0.616)--(8.805,0.616)--(8.805,0.639)--(8.697,0.639)--cycle;
\draw[gp path] (8.697,0.616)--(8.697,0.638)--(8.804,0.638)--(8.804,0.616)--cycle;
\gpfill{rgb color={\colora}} (8.850,0.616)--(8.957,0.616)--(8.957,0.633)--(8.850,0.633)--cycle;
\draw[gp path] (8.850,0.616)--(8.850,0.632)--(8.956,0.632)--(8.956,0.616)--cycle;
\gpfill{rgb color={\colora}} (9.002,0.616)--(9.109,0.616)--(9.109,0.629)--(9.002,0.629)--cycle;
\draw[gp path] (9.002,0.616)--(9.002,0.628)--(9.108,0.628)--(9.108,0.616)--cycle;
\gpfill{rgb color={\colora}} (9.154,0.616)--(9.262,0.616)--(9.262,0.625)--(9.154,0.625)--cycle;
\draw[gp path] (9.154,0.616)--(9.154,0.624)--(9.261,0.624)--(9.261,0.616)--cycle;
\gpfill{rgb color={\colora}} (9.306,0.616)--(9.414,0.616)--(9.414,0.622)--(9.306,0.622)--cycle;
\draw[gp path] (9.306,0.616)--(9.306,0.621)--(9.413,0.621)--(9.413,0.616)--cycle;
\gpfill{rgb color={\colora}} (9.459,0.616)--(9.566,0.616)--(9.566,0.620)--(9.459,0.620)--cycle;
\draw[gp path] (9.459,0.616)--(9.459,0.619)--(9.565,0.619)--(9.565,0.616)--cycle;
\gpfill{rgb color={\colora}} (9.611,0.616)--(9.719,0.616)--(9.719,0.619)--(9.611,0.619)--cycle;
\draw[gp path] (9.611,0.616)--(9.611,0.618)--(9.718,0.618)--(9.718,0.616)--cycle;
\gpfill{rgb color={\colora}} (9.763,0.616)--(9.871,0.616)--(9.871,0.618)--(9.763,0.618)--cycle;
\draw[gp path] (9.763,0.616)--(9.763,0.617)--(9.870,0.617)--(9.870,0.616)--cycle;
\gpfill{rgb color={\colora}} (9.915,0.616)--(10.023,0.616)--(10.023,0.617)--(9.915,0.617)--cycle;
\draw[gp path] (9.915,0.616)--(10.022,0.616)--cycle;
\gpfill{rgb color={\colora}} (10.068,0.616)--(10.175,0.616)--(10.175,0.617)--(10.068,0.617)--cycle;
\draw[gp path] (10.068,0.616)--(10.174,0.616)--cycle;
\gpfill{rgb color={\colora}} (10.220,0.616)--(10.328,0.616)--(10.328,0.617)--(10.220,0.617)--cycle;
\draw[gp path] (10.220,0.616)--(10.327,0.616)--cycle;
\gpfill{rgb color={\colora}} (10.372,0.616)--(10.480,0.616)--(10.480,0.617)--(10.372,0.617)--cycle;
\draw[gp path] (10.372,0.616)--(10.479,0.616)--cycle;
\gpfill{rgb color={\colora}} (10.524,0.616)--(10.632,0.616)--(10.632,0.617)--(10.524,0.617)--cycle;
\draw[gp path] (10.524,0.616)--(10.631,0.616)--cycle;
\gpcolor{color=gp lt color border}
\draw[gp path] (2.965,8.381)--(2.965,0.616)--(10.730,0.616)--(10.730,8.381)--cycle;
\gpdefrectangularnode{gp plot 1}{\pgfpoint{2.965cm}{0.616cm}}{\pgfpoint{10.730cm}{8.381cm}}
\end{tikzpicture}\!\!\!\!
\begin{tikzpicture}[scale=0.55,gnuplot]
\path (1.000,0.000) rectangle (12.500,8.750);
\gpcolor{color=gp lt color border}
\gpsetlinetype{gp lt border}
\gpsetdashtype{gp dt solid}
\gpsetlinewidth{1.00}
\draw[gp path] (2.965,0.616)--(3.145,0.616);
\draw[gp path] (10.730,0.616)--(10.550,0.616);
\node[gp node right] at (2.781,0.616) {\scriptsize $0$};
\draw[gp path] (2.965,1.733)--(3.145,1.733);
\draw[gp path] (10.730,1.733)--(10.550,1.733);
\node[gp node right] at (2.781,1.733) {\scriptsize $0.2\!\cdot\!10^{6}\!\!\!$};
\draw[gp path] (2.965,2.851)--(3.145,2.851);
\draw[gp path] (10.730,2.851)--(10.550,2.851);
\node[gp node right] at (2.781,2.851) {\scriptsize $0.4\!\cdot\!10^{6}\!\!\!$};
\draw[gp path] (2.965,3.968)--(3.145,3.968);
\draw[gp path] (10.730,3.968)--(10.550,3.968);
\node[gp node right] at (2.781,3.968) {\scriptsize $0.6\!\cdot\!10^{6}\!\!\!$};
\draw[gp path] (2.965,5.085)--(3.145,5.085);
\draw[gp path] (10.730,5.085)--(10.550,5.085);
\node[gp node right] at (2.781,5.085) {\scriptsize $0.8\!\cdot\!10^{6}\!\!\!$};
\draw[gp path] (2.965,6.202)--(3.145,6.202);
\draw[gp path] (10.730,6.202)--(10.550,6.202);
\node[gp node right] at (2.781,6.202) {\scriptsize $1.0\!\cdot\!10^{6}\!\!\!$};
\draw[gp path] (2.965,7.320)--(3.145,7.320);
\draw[gp path] (10.730,7.320)--(10.550,7.320);
\node[gp node right] at (2.781,7.320) {\scriptsize $1.2\!\cdot\!10^{6}\!\!\!$};
\draw[gp path] (3.396,0.616)--(3.396,0.796);
\draw[gp path] (3.396,8.381)--(3.396,8.201);
\node[gp node center] at (3.396,0.308) {\scriptsize $3400$};
\draw[gp path] (5.122,0.616)--(5.122,0.796);
\draw[gp path] (5.122,8.381)--(5.122,8.201);
\node[gp node center] at (5.122,0.308) {\scriptsize $3600$};
\draw[gp path] (6.848,0.616)--(6.848,0.796);
\draw[gp path] (6.848,8.381)--(6.848,8.201);
\node[gp node center] at (6.848,0.308) {\scriptsize $3800$};
\draw[gp path] (8.573,0.616)--(8.573,0.796);
\draw[gp path] (8.573,8.381)--(8.573,8.201);
\node[gp node center] at (8.573,0.308) {\scriptsize $4000$};
\draw[gp path] (10.299,0.616)--(10.299,0.796);
\draw[gp path] (10.299,8.381)--(10.299,8.201);
\node[gp node center] at (10.299,0.308) {\scriptsize $4200$};
\draw[gp path] (2.965,8.381)--(2.965,0.616)--(10.730,0.616)--(10.730,8.381)--cycle;
\gpfill{rgb color={0.580,0.000,0.827}} (2.987,0.616)--(3.117,0.616)--(3.117,0.628)--(2.987,0.628)--cycle;
\gpcolor{rgb color={\colorb}}
\draw[gp path] (2.987,0.616)--(2.987,0.627)--(3.116,0.627)--(3.116,0.616)--cycle;
\gpfill{rgb color={\colorb}} (3.159,0.616)--(3.290,0.616)--(3.290,0.630)--(3.159,0.630)--cycle;
\draw[gp path] (3.159,0.616)--(3.159,0.629)--(3.289,0.629)--(3.289,0.616)--cycle;
\gpfill{rgb color={\colorb}} (3.332,0.616)--(3.462,0.616)--(3.462,0.633)--(3.332,0.633)--cycle;
\draw[gp path] (3.332,0.616)--(3.332,0.632)--(3.461,0.632)--(3.461,0.616)--cycle;
\gpfill{rgb color={\colorb}} (3.504,0.616)--(3.635,0.616)--(3.635,0.637)--(3.504,0.637)--cycle;
\draw[gp path] (3.504,0.616)--(3.504,0.636)--(3.634,0.636)--(3.634,0.616)--cycle;
\gpfill{rgb color={\colorb}} (3.677,0.616)--(3.807,0.616)--(3.807,0.642)--(3.677,0.642)--cycle;
\draw[gp path] (3.677,0.616)--(3.677,0.641)--(3.806,0.641)--(3.806,0.616)--cycle;
\gpfill{rgb color={\colorb}} (3.849,0.616)--(3.980,0.616)--(3.980,0.647)--(3.849,0.647)--cycle;
\draw[gp path] (3.849,0.616)--(3.849,0.646)--(3.979,0.646)--(3.979,0.616)--cycle;
\gpfill{rgb color={\colorb}} (4.022,0.616)--(4.152,0.616)--(4.152,0.653)--(4.022,0.653)--cycle;
\draw[gp path] (4.022,0.616)--(4.022,0.652)--(4.151,0.652)--(4.151,0.616)--cycle;
\gpfill{rgb color={\colorb}} (4.194,0.616)--(4.325,0.616)--(4.325,0.659)--(4.194,0.659)--cycle;
\draw[gp path] (4.194,0.616)--(4.194,0.658)--(4.324,0.658)--(4.324,0.616)--cycle;
\gpfill{rgb color={\colorb}} (4.367,0.616)--(4.497,0.616)--(4.497,0.667)--(4.367,0.667)--cycle;
\draw[gp path] (4.367,0.616)--(4.367,0.666)--(4.496,0.666)--(4.496,0.616)--cycle;
\gpfill{rgb color={\colorb}} (4.540,0.616)--(4.670,0.616)--(4.670,0.679)--(4.540,0.679)--cycle;
\draw[gp path] (4.540,0.616)--(4.540,0.678)--(4.669,0.678)--(4.669,0.616)--cycle;
\gpfill{rgb color={\colorb}} (4.712,0.616)--(4.843,0.616)--(4.843,0.697)--(4.712,0.697)--cycle;
\draw[gp path] (4.712,0.616)--(4.712,0.696)--(4.842,0.696)--(4.842,0.616)--cycle;
\gpfill{rgb color={\colorb}} (4.885,0.616)--(5.015,0.616)--(5.015,0.734)--(4.885,0.734)--cycle;
\draw[gp path] (4.885,0.616)--(4.885,0.733)--(5.014,0.733)--(5.014,0.616)--cycle;
\gpfill{rgb color={\colorb}} (5.057,0.616)--(5.188,0.616)--(5.188,0.839)--(5.057,0.839)--cycle;
\draw[gp path] (5.057,0.616)--(5.057,0.838)--(5.187,0.838)--(5.187,0.616)--cycle;
\gpfill{rgb color={\colorb}} (5.230,0.616)--(5.360,0.616)--(5.360,1.545)--(5.230,1.545)--cycle;
\draw[gp path] (5.230,0.616)--(5.230,1.544)--(5.359,1.544)--(5.359,0.616)--cycle;
\gpfill{rgb color={\colorb}} (5.402,0.616)--(5.533,0.616)--(5.533,2.935)--(5.402,2.935)--cycle;
\draw[gp path] (5.402,0.616)--(5.402,2.934)--(5.532,2.934)--(5.532,0.616)--cycle;
\gpfill{rgb color={\colorb}} (5.575,0.616)--(5.705,0.616)--(5.705,5.030)--(5.575,5.030)--cycle;
\draw[gp path] (5.575,0.616)--(5.575,5.029)--(5.704,5.029)--(5.704,0.616)--cycle;
\gpfill{rgb color={\colorb}} (5.747,0.616)--(5.878,0.616)--(5.878,6.781)--(5.747,6.781)--cycle;
\draw[gp path] (5.747,0.616)--(5.747,6.780)--(5.877,6.780)--(5.877,0.616)--cycle;
\gpfill{rgb color={\colorb}} (5.920,0.616)--(6.050,0.616)--(6.050,7.489)--(5.920,7.489)--cycle;
\draw[gp path] (5.920,0.616)--(5.920,7.488)--(6.049,7.488)--(6.049,0.616)--cycle;
\gpfill{rgb color={\colorb}} (6.093,0.616)--(6.223,0.616)--(6.223,7.998)--(6.093,7.998)--cycle;
\draw[gp path] (6.093,0.616)--(6.093,7.997)--(6.222,7.997)--(6.222,0.616)--cycle;
\gpfill{rgb color={\colorb}} (6.265,0.616)--(6.396,0.616)--(6.396,7.365)--(6.265,7.365)--cycle;
\draw[gp path] (6.265,0.616)--(6.265,7.364)--(6.395,7.364)--(6.395,0.616)--cycle;
\gpfill{rgb color={\colorb}} (6.438,0.616)--(6.568,0.616)--(6.568,5.663)--(6.438,5.663)--cycle;
\draw[gp path] (6.438,0.616)--(6.438,5.662)--(6.567,5.662)--(6.567,0.616)--cycle;
\gpfill{rgb color={\colorb}} (6.610,0.616)--(6.741,0.616)--(6.741,4.159)--(6.610,4.159)--cycle;
\draw[gp path] (6.610,0.616)--(6.610,4.158)--(6.740,4.158)--(6.740,0.616)--cycle;
\gpfill{rgb color={\colorb}} (6.783,0.616)--(6.913,0.616)--(6.913,3.562)--(6.783,3.562)--cycle;
\draw[gp path] (6.783,0.616)--(6.783,3.561)--(6.912,3.561)--(6.912,0.616)--cycle;
\gpfill{rgb color={\colorb}} (6.955,0.616)--(7.086,0.616)--(7.086,3.140)--(6.955,3.140)--cycle;
\draw[gp path] (6.955,0.616)--(6.955,3.139)--(7.085,3.139)--(7.085,0.616)--cycle;
\gpfill{rgb color={\colorb}} (7.128,0.616)--(7.258,0.616)--(7.258,2.551)--(7.128,2.551)--cycle;
\draw[gp path] (7.128,0.616)--(7.128,2.550)--(7.257,2.550)--(7.257,0.616)--cycle;
\gpfill{rgb color={\colorb}} (7.300,0.616)--(7.431,0.616)--(7.431,1.992)--(7.300,1.992)--cycle;
\draw[gp path] (7.300,0.616)--(7.300,1.991)--(7.430,1.991)--(7.430,0.616)--cycle;
\gpfill{rgb color={\colorb}} (7.473,0.616)--(7.603,0.616)--(7.603,1.667)--(7.473,1.667)--cycle;
\draw[gp path] (7.473,0.616)--(7.473,1.666)--(7.602,1.666)--(7.602,0.616)--cycle;
\gpfill{rgb color={\colorb}} (7.646,0.616)--(7.776,0.616)--(7.776,1.489)--(7.646,1.489)--cycle;
\draw[gp path] (7.646,0.616)--(7.646,1.488)--(7.775,1.488)--(7.775,0.616)--cycle;
\gpfill{rgb color={\colorb}} (7.818,0.616)--(7.949,0.616)--(7.949,1.306)--(7.818,1.306)--cycle;
\draw[gp path] (7.818,0.616)--(7.818,1.305)--(7.948,1.305)--(7.948,0.616)--cycle;
\gpfill{rgb color={\colorb}} (7.991,0.616)--(8.121,0.616)--(8.121,1.146)--(7.991,1.146)--cycle;
\draw[gp path] (7.991,0.616)--(7.991,1.145)--(8.120,1.145)--(8.120,0.616)--cycle;
\gpfill{rgb color={\colorb}} (8.163,0.616)--(8.294,0.616)--(8.294,1.047)--(8.163,1.047)--cycle;
\draw[gp path] (8.163,0.616)--(8.163,1.046)--(8.293,1.046)--(8.293,0.616)--cycle;
\gpfill{rgb color={\colorb}} (8.336,0.616)--(8.466,0.616)--(8.466,0.961)--(8.336,0.961)--cycle;
\draw[gp path] (8.336,0.616)--(8.336,0.960)--(8.465,0.960)--(8.465,0.616)--cycle;
\gpfill{rgb color={\colorb}} (8.508,0.616)--(8.639,0.616)--(8.639,0.874)--(8.508,0.874)--cycle;
\draw[gp path] (8.508,0.616)--(8.508,0.873)--(8.638,0.873)--(8.638,0.616)--cycle;
\gpfill{rgb color={\colorb}} (8.681,0.616)--(8.811,0.616)--(8.811,0.811)--(8.681,0.811)--cycle;
\draw[gp path] (8.681,0.616)--(8.681,0.810)--(8.810,0.810)--(8.810,0.616)--cycle;
\gpfill{rgb color={\colorb}} (8.853,0.616)--(8.984,0.616)--(8.984,0.759)--(8.853,0.759)--cycle;
\draw[gp path] (8.853,0.616)--(8.853,0.758)--(8.983,0.758)--(8.983,0.616)--cycle;
\gpfill{rgb color={\colorb}} (9.026,0.616)--(9.156,0.616)--(9.156,0.714)--(9.026,0.714)--cycle;
\draw[gp path] (9.026,0.616)--(9.026,0.713)--(9.155,0.713)--(9.155,0.616)--cycle;
\gpfill{rgb color={\colorb}} (9.199,0.616)--(9.329,0.616)--(9.329,0.680)--(9.199,0.680)--cycle;
\draw[gp path] (9.199,0.616)--(9.199,0.679)--(9.328,0.679)--(9.328,0.616)--cycle;
\gpfill{rgb color={\colorb}} (9.371,0.616)--(9.502,0.616)--(9.502,0.659)--(9.371,0.659)--cycle;
\draw[gp path] (9.371,0.616)--(9.371,0.658)--(9.501,0.658)--(9.501,0.616)--cycle;
\gpfill{rgb color={\colorb}} (9.544,0.616)--(9.674,0.616)--(9.674,0.643)--(9.544,0.643)--cycle;
\draw[gp path] (9.544,0.616)--(9.544,0.642)--(9.673,0.642)--(9.673,0.616)--cycle;
\gpfill{rgb color={\colorb}} (9.716,0.616)--(9.847,0.616)--(9.847,0.631)--(9.716,0.631)--cycle;
\draw[gp path] (9.716,0.616)--(9.716,0.630)--(9.846,0.630)--(9.846,0.616)--cycle;
\gpfill{rgb color={\colorb}} (9.889,0.616)--(10.019,0.616)--(10.019,0.625)--(9.889,0.625)--cycle;
\draw[gp path] (9.889,0.616)--(9.889,0.624)--(10.018,0.624)--(10.018,0.616)--cycle;
\gpfill{rgb color={\colorb}} (10.061,0.616)--(10.192,0.616)--(10.192,0.621)--(10.061,0.621)--cycle;
\draw[gp path] (10.061,0.616)--(10.061,0.620)--(10.191,0.620)--(10.191,0.616)--cycle;
\gpfill{rgb color={\colorb}} (10.234,0.616)--(10.364,0.616)--(10.364,0.619)--(10.234,0.619)--cycle;
\draw[gp path] (10.234,0.616)--(10.234,0.618)--(10.363,0.618)--(10.363,0.616)--cycle;
\gpfill{rgb color={\colorb}} (10.406,0.616)--(10.537,0.616)--(10.537,0.618)--(10.406,0.618)--cycle;
\draw[gp path] (10.406,0.616)--(10.406,0.617)--(10.536,0.617)--(10.536,0.616)--cycle;
\gpfill{rgb color={\colorb}} (10.579,0.616)--(10.709,0.616)--(10.709,0.617)--(10.579,0.617)--cycle;
\draw[gp path] (10.579,0.616)--(10.708,0.616)--cycle;
\gpcolor{color=gp lt color border}
\draw[gp path] (2.965,8.381)--(2.965,0.616)--(10.730,0.616)--(10.730,8.381)--cycle;
\gpdefrectangularnode{gp plot 1}{\pgfpoint{2.965cm}{0.616cm}}{\pgfpoint{10.730cm}{8.381cm}}
\end{tikzpicture}\!\!\!\!
\begin{tikzpicture}[scale=0.55,gnuplot]
\path (1.000,0.000) rectangle (12.500,8.750);
\gpcolor{color=gp lt color border}
\gpsetlinetype{gp lt border}
\gpsetdashtype{gp dt solid}
\gpsetlinewidth{1.00}
\draw[gp path] (2.965,0.616)--(3.145,0.616);
\draw[gp path] (10.730,0.616)--(10.550,0.616);
\node[gp node right] at (2.781,0.616) {\scriptsize $0$};
\draw[gp path] (2.965,1.733)--(3.145,1.733);
\draw[gp path] (10.730,1.733)--(10.550,1.733);
\node[gp node right] at (2.781,1.733) {\scriptsize $0.2\!\cdot\!10^{6}\!\!\!$};
\draw[gp path] (2.965,2.851)--(3.145,2.851);
\draw[gp path] (10.730,2.851)--(10.550,2.851);
\node[gp node right] at (2.781,2.851) {\scriptsize $0.4\!\cdot\!10^{6}\!\!\!$};
\draw[gp path] (2.965,3.968)--(3.145,3.968);
\draw[gp path] (10.730,3.968)--(10.550,3.968);
\node[gp node right] at (2.781,3.968) {\scriptsize $0.6\!\cdot\!10^{6}\!\!\!$};
\draw[gp path] (2.965,5.085)--(3.145,5.085);
\draw[gp path] (10.730,5.085)--(10.550,5.085);
\node[gp node right] at (2.781,5.085) {\scriptsize $0.8\!\cdot\!10^{6}\!\!\!$};
\draw[gp path] (2.965,6.202)--(3.145,6.202);
\draw[gp path] (10.730,6.202)--(10.550,6.202);
\node[gp node right] at (2.781,6.202) {\scriptsize $1.0\!\cdot\!10^{6}\!\!\!$};
\draw[gp path] (2.965,7.320)--(3.145,7.320);
\draw[gp path] (10.730,7.320)--(10.550,7.320);
\node[gp node right] at (2.781,7.320) {\scriptsize $1.2\!\cdot\!10^{6}\!\!\!$};
\draw[gp path] (2.965,0.616)--(2.965,0.796);
\draw[gp path] (2.965,8.381)--(2.965,8.201);
\node[gp node center] at (2.965,0.308) {\scriptsize $0$};
\draw[gp path] (3.936,0.616)--(3.936,0.796);
\draw[gp path] (3.936,8.381)--(3.936,8.201);
\node[gp node center] at (3.936,0.308) {\scriptsize $20$};
\draw[gp path] (4.906,0.616)--(4.906,0.796);
\draw[gp path] (4.906,8.381)--(4.906,8.201);
\node[gp node center] at (4.906,0.308) {\scriptsize $40$};
\draw[gp path] (5.877,0.616)--(5.877,0.796);
\draw[gp path] (5.877,8.381)--(5.877,8.201);
\node[gp node center] at (5.877,0.308) {\scriptsize $60$};
\draw[gp path] (6.848,0.616)--(6.848,0.796);
\draw[gp path] (6.848,8.381)--(6.848,8.201);
\node[gp node center] at (6.848,0.308) {\scriptsize $80$};
\draw[gp path] (7.818,0.616)--(7.818,0.796);
\draw[gp path] (7.818,8.381)--(7.818,8.201);
\node[gp node center] at (7.818,0.308) {\scriptsize $100$};
\draw[gp path] (8.789,0.616)--(8.789,0.796);
\draw[gp path] (8.789,8.381)--(8.789,8.201);
\node[gp node center] at (8.789,0.308) {\scriptsize $120$};
\draw[gp path] (9.759,0.616)--(9.759,0.796);
\draw[gp path] (9.759,8.381)--(9.759,8.201);
\node[gp node center] at (9.759,0.308) {\scriptsize $140$};
\draw[gp path] (10.730,0.616)--(10.730,0.796);
\draw[gp path] (10.730,8.381)--(10.730,8.201);
\node[gp node center] at (10.730,0.308) {\scriptsize $160$};
\draw[gp path] (2.965,8.381)--(2.965,0.616)--(10.730,0.616)--(10.730,8.381)--cycle;
\gpfill{rgb color={\colorc}} (3.001,0.616)--(3.172,0.616)--(3.172,1.440)--(3.001,1.440)--cycle;
\gpcolor{rgb color={\colorc}}
\draw[gp path] (3.001,0.616)--(3.001,1.439)--(3.171,1.439)--(3.171,0.616)--cycle;
\gpfill{rgb color={\colorc}} (3.244,0.616)--(3.415,0.616)--(3.415,2.424)--(3.244,2.424)--cycle;
\draw[gp path] (3.244,0.616)--(3.244,2.423)--(3.414,2.423)--(3.414,0.616)--cycle;
\gpfill{rgb color={\colorc}} (3.487,0.616)--(3.658,0.616)--(3.658,3.572)--(3.487,3.572)--cycle;
\draw[gp path] (3.487,0.616)--(3.487,3.571)--(3.657,3.571)--(3.657,0.616)--cycle;
\gpfill{rgb color={\colorc}} (3.729,0.616)--(3.900,0.616)--(3.900,4.428)--(3.729,4.428)--cycle;
\draw[gp path] (3.729,0.616)--(3.729,4.427)--(3.899,4.427)--(3.899,0.616)--cycle;
\gpfill{rgb color={\colorc}} (3.972,0.616)--(4.143,0.616)--(4.143,4.736)--(3.972,4.736)--cycle;
\draw[gp path] (3.972,0.616)--(3.972,4.735)--(4.142,4.735)--(4.142,0.616)--cycle;
\gpfill{rgb color={\colorc}} (4.215,0.616)--(4.386,0.616)--(4.386,4.726)--(4.215,4.726)--cycle;
\draw[gp path] (4.215,0.616)--(4.215,4.725)--(4.385,4.725)--(4.385,0.616)--cycle;
\gpfill{rgb color={\colorc}} (4.457,0.616)--(4.628,0.616)--(4.628,4.562)--(4.457,4.562)--cycle;
\draw[gp path] (4.457,0.616)--(4.457,4.561)--(4.627,4.561)--(4.627,0.616)--cycle;
\gpfill{rgb color={\colorc}} (4.700,0.616)--(4.871,0.616)--(4.871,4.362)--(4.700,4.362)--cycle;
\draw[gp path] (4.700,0.616)--(4.700,4.361)--(4.870,4.361)--(4.870,0.616)--cycle;
\gpfill{rgb color={\colorc}} (4.943,0.616)--(5.114,0.616)--(5.114,4.121)--(4.943,4.121)--cycle;
\draw[gp path] (4.943,0.616)--(4.943,4.120)--(5.113,4.120)--(5.113,0.616)--cycle;
\gpfill{rgb color={\colorc}} (5.185,0.616)--(5.356,0.616)--(5.356,3.869)--(5.185,3.869)--cycle;
\draw[gp path] (5.185,0.616)--(5.185,3.868)--(5.355,3.868)--(5.355,0.616)--cycle;
\gpfill{rgb color={\colorc}} (5.428,0.616)--(5.599,0.616)--(5.599,3.602)--(5.428,3.602)--cycle;
\draw[gp path] (5.428,0.616)--(5.428,3.601)--(5.598,3.601)--(5.598,0.616)--cycle;
\gpfill{rgb color={\colorc}} (5.671,0.616)--(5.841,0.616)--(5.841,3.334)--(5.671,3.334)--cycle;
\draw[gp path] (5.671,0.616)--(5.671,3.333)--(5.840,3.333)--(5.840,0.616)--cycle;
\gpfill{rgb color={\colorc}} (5.913,0.616)--(6.084,0.616)--(6.084,3.072)--(5.913,3.072)--cycle;
\draw[gp path] (5.913,0.616)--(5.913,3.071)--(6.083,3.071)--(6.083,0.616)--cycle;
\gpfill{rgb color={\colorc}} (6.156,0.616)--(6.327,0.616)--(6.327,2.821)--(6.156,2.821)--cycle;
\draw[gp path] (6.156,0.616)--(6.156,2.820)--(6.326,2.820)--(6.326,0.616)--cycle;
\gpfill{rgb color={\colorc}} (6.399,0.616)--(6.569,0.616)--(6.569,2.566)--(6.399,2.566)--cycle;
\draw[gp path] (6.399,0.616)--(6.399,2.565)--(6.568,2.565)--(6.568,0.616)--cycle;
\gpfill{rgb color={\colorc}} (6.641,0.616)--(6.812,0.616)--(6.812,2.299)--(6.641,2.299)--cycle;
\draw[gp path] (6.641,0.616)--(6.641,2.298)--(6.811,2.298)--(6.811,0.616)--cycle;
\gpfill{rgb color={\colorc}} (6.884,0.616)--(7.055,0.616)--(7.055,2.030)--(6.884,2.030)--cycle;
\draw[gp path] (6.884,0.616)--(6.884,2.029)--(7.054,2.029)--(7.054,0.616)--cycle;
\gpfill{rgb color={\colorc}} (7.127,0.616)--(7.297,0.616)--(7.297,1.754)--(7.127,1.754)--cycle;
\draw[gp path] (7.127,0.616)--(7.127,1.753)--(7.296,1.753)--(7.296,0.616)--cycle;
\gpfill{rgb color={\colorc}} (7.369,0.616)--(7.540,0.616)--(7.540,1.508)--(7.369,1.508)--cycle;
\draw[gp path] (7.369,0.616)--(7.369,1.507)--(7.539,1.507)--(7.539,0.616)--cycle;
\gpfill{rgb color={\colorc}} (7.612,0.616)--(7.783,0.616)--(7.783,1.316)--(7.612,1.316)--cycle;
\draw[gp path] (7.612,0.616)--(7.612,1.315)--(7.782,1.315)--(7.782,0.616)--cycle;
\gpfill{rgb color={\colorc}} (7.855,0.616)--(8.025,0.616)--(8.025,1.166)--(7.855,1.166)--cycle;
\draw[gp path] (7.855,0.616)--(7.855,1.165)--(8.024,1.165)--(8.024,0.616)--cycle;
\gpfill{rgb color={\colorc}} (8.097,0.616)--(8.268,0.616)--(8.268,1.051)--(8.097,1.051)--cycle;
\draw[gp path] (8.097,0.616)--(8.097,1.050)--(8.267,1.050)--(8.267,0.616)--cycle;
\gpfill{rgb color={\colorc}} (8.340,0.616)--(8.511,0.616)--(8.511,0.964)--(8.340,0.964)--cycle;
\draw[gp path] (8.340,0.616)--(8.340,0.963)--(8.510,0.963)--(8.510,0.616)--cycle;
\gpfill{rgb color={\colorc}} (8.582,0.616)--(8.753,0.616)--(8.753,0.897)--(8.582,0.897)--cycle;
\draw[gp path] (8.582,0.616)--(8.582,0.896)--(8.752,0.896)--(8.752,0.616)--cycle;
\gpfill{rgb color={\colorc}} (8.825,0.616)--(8.996,0.616)--(8.996,0.844)--(8.825,0.844)--cycle;
\draw[gp path] (8.825,0.616)--(8.825,0.843)--(8.995,0.843)--(8.995,0.616)--cycle;
\gpfill{rgb color={\colorc}} (9.068,0.616)--(9.239,0.616)--(9.239,0.800)--(9.068,0.800)--cycle;
\draw[gp path] (9.068,0.616)--(9.068,0.799)--(9.238,0.799)--(9.238,0.616)--cycle;
\gpfill{rgb color={\colorc}} (9.310,0.616)--(9.481,0.616)--(9.481,0.757)--(9.310,0.757)--cycle;
\draw[gp path] (9.310,0.616)--(9.310,0.756)--(9.480,0.756)--(9.480,0.616)--cycle;
\gpfill{rgb color={\colorc}} (9.553,0.616)--(9.724,0.616)--(9.724,0.710)--(9.553,0.710)--cycle;
\draw[gp path] (9.553,0.616)--(9.553,0.709)--(9.723,0.709)--(9.723,0.616)--cycle;
\gpfill{rgb color={\colorc}} (9.796,0.616)--(9.967,0.616)--(9.967,0.673)--(9.796,0.673)--cycle;
\draw[gp path] (9.796,0.616)--(9.796,0.672)--(9.966,0.672)--(9.966,0.616)--cycle;
\gpfill{rgb color={\colorc}} (10.038,0.616)--(10.209,0.616)--(10.209,0.654)--(10.038,0.654)--cycle;
\draw[gp path] (10.038,0.616)--(10.038,0.653)--(10.208,0.653)--(10.208,0.616)--cycle;
\gpfill{rgb color={\colorc}} (10.281,0.616)--(10.452,0.616)--(10.452,0.640)--(10.281,0.640)--cycle;
\draw[gp path] (10.281,0.616)--(10.281,0.639)--(10.451,0.639)--(10.451,0.616)--cycle;
\gpfill{rgb color={\colorc}} (10.524,0.616)--(10.695,0.616)--(10.695,0.628)--(10.524,0.628)--cycle;
\draw[gp path] (10.524,0.616)--(10.524,0.627)--(10.694,0.627)--(10.694,0.616)--cycle;
\gpcolor{color=gp lt color border}
\draw[gp path] (2.965,8.381)--(2.965,0.616)--(10.730,0.616)--(10.730,8.381)--cycle;
\gpdefrectangularnode{gp plot 1}{\pgfpoint{2.965cm}{0.616cm}}{\pgfpoint{10.730cm}{8.381cm}}
\end{tikzpicture}

\caption{Histograms showing statistics regarding the sizes of cubes (left, ranging from $13$ to $62$), the number of binary clauses in the formula under that cube (middle), 
and the time to solve the subformulas (right, in seconds).}\
\label{fig:histograms}
\end{figure*}
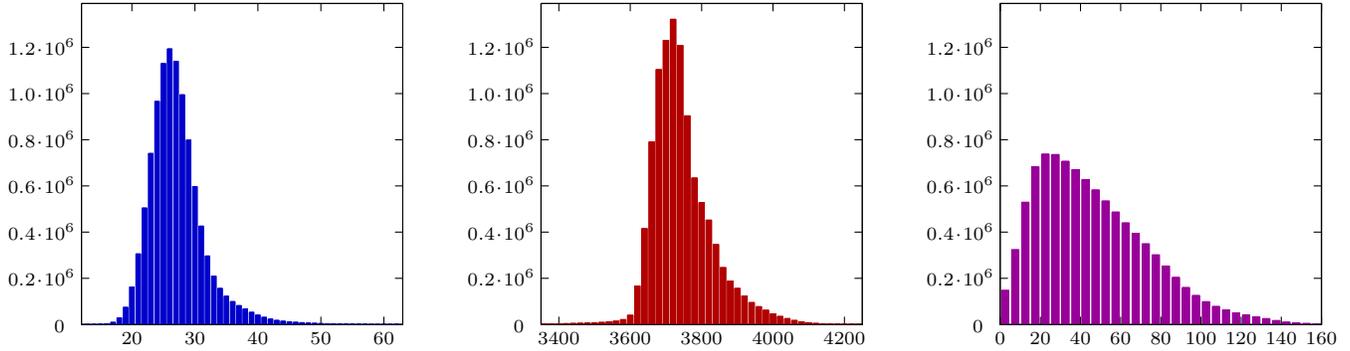

%

\subsection{Solving Subproblems}

Our top-level partition of $R^{5}_{160}$ and $R^{5}_{161}$, denoted by $P^5$, consists of $10\,330\,615$ cubes after partition balancing.
Figure~\ref{fig:histograms} (left) shows a histogram of the size of the cubes in $P^5$. The smallest cube has size $13$ while the largest
cube has length $62$, showing that the binary tree associated with the cubes is quite unbalanced. Notice that the size of most cubes is in the
range from $20$ to $40$, which is a large interval. 

Figure~\ref{fig:histograms} (middle) shows a histogram of the number
of binary clauses in subproblems, i.e., the resulting formulas after applying the cubes. Notice that the interval here is small: Most subproblems have between $3650$ and $3850$ binary clauses.
As stated earlier, the number of binary clauses is a useful rough measure for the hardness of subproblems.

For each cube $\alpha \in P^5$, we solved\footnote{The tools and proof parts presented in this paper are available at \url{https://www.cs.utexas.edu/~marijn/Schur/}.}
the problem $R^{5}_{160} \land \alpha$ using our cube-and-conquer solver
consisting of a modified version of {\sc March$\_$cu}~\cite{HKWB11} as look-ahead (cube) solver and {\sc Glucose} 3.0~\cite{glucose} as CDCL (conquer) solver.
The cube solver modifications consist of integrating the presented decision heuristic 
and replacing the cutoff procedure by the presented down factor mechanism.
In case $R^{5}_{160} \land \alpha$ was unsatisfiable, we stored the proof of unsatisfiability, which is also a proof of 
unsatisfiability of $R^{5}_{161} \land \alpha$. There were only 961 cubes $\alpha \in P_5$ for which $R^{5}_{160} \land \alpha$
turned out to be satisfiable. For those cubes, we computed the proof of unsatisfiability of $R^{5}_{161} \land \alpha$.
These proofs together form the {\em implication proof}.

We solved the subproblems on the Lonestar $5$ cluster of the Texas Advanced Computing Center (TACC). Each compute node
consists of a Xeon E5-2690 v3 (Haswell) chip with $24$ cores running on $2.6$ GHz. Hyper-threading was enabled, resulting in $48$ logical CPUs per node.
We ran the experiments on $50$ nodes in parallel, resulting in running $2400$ copies of our cube-and-conquer solver 
in parallel. 
The total runtime was roughly $27\,600$ CPU hours for the partition phase and roughly $95\,600$ CPU hours for the conquer phase.
The total costs to compute Schur Number Five was just over $14$ CPU years, but less than three days in wall-clock time on the Lonestar $5$ cluster.
Figure~\ref{fig:histograms} (right) shows a histogram of the runtimes (rounding times to the nearest $5$ seconds). A large fraction of the subproblems can be solved within 
$20$ to $40$ seconds. Most subproblems are solvable within two minutes and few are somewhat harder.
The subproblems were partitioned into a total of $65$ billion cubes and the number of conflict clauses added in the conquer phase was $11$ trillion.  

The computation of Schur Number Five and the Pythagorean triples problem differ in the balance between the partition phase and the conquer phase. 
For Schur Number Five, almost $78\%$ of the computation was devoted to the conquer phase, while for the Pythagorean triples problem this was only $38\%$. 
This difference can be explained as follows: For both problems, a heuristic was chosen to continue splitting until the total runtime would start to increase
(based on the solving time of randomly selected subproblems).
In the case of Schur Number Five, this point is reached earlier. The Pythagorean triples problem was solved on the older Stampede cluster of TACC, 
which hinders a clean runtime comparison. We estimate that the conquer phase of solving Schur Number Five required
about ten times more computation resources than solving the Pythagorean triples problem.

\section{No Backbone, but Backdoors}

The {\em backbone} of a CNF formula is the set of literals that are assigned to true in
all satisfying total assignments. Many formulas that encode the existence of extreme certificates of problems in Ramsey Theory~\cite{RamseyTheory}, 
such as the van der Waerden numbers~\cite{KP08} and the Pythagorean triples problem~\cite{ptn},
have large backbones after symmetry breaking---even if the number of satisfying total assignments is enormous. 
However, the backbone of $R^{5}_{160}$ consists only of the literals that are assigned by the symmetry-breaking predicates.

The lack of a substantial backbone suggests that there may exist a symmetry that is not broken. It turns out that 
palindromic Schur number problems have certificate symmetry $\sigma_{\cdot p}$ that
maps each number $i$ onto $i \cdot p \pmod{n+1}$ with $n$ being the size of the certificate and $p$ any number
that is relatively prime to $n+1$~\cite{Palin}. However, $\sigma_{\cdot p}$ is not a certificate symmetry of 
classic Schur number problems. The size of the backbone can therefore be explained by 
the equivalence $S(5) = S_\mathrm{pd}(5) = 160$ and the certificate symmetry $\sigma_{\cdot p}$ of 
palindromic Schur number problems.


Although the backbone of $R^{5}_{160}$ is small, we observed that there are several backdoors to large clusters of solutions.
A {\em backdoor} of a CNF formula $F$ is a partial assignment $\beta$ such that $F \land \beta$ 
can be solved efficiently (in polynomial time) using a given algorithm~\cite{backdoor}. 
We selected subsumption elimination (SE)~\cite{BVE} followed by blocked clause elimination (BCE)~\cite{BCE} as the basis for backdoors of $R^{5}_{160}$. 
Both SE and BCE are confluent and run in polynomial time. BCE removes blocked clauses until fixpoint. 
BCE solves a given formula if and only if the fixpoint is the empty formula, which is trivially satisfiable. 

We computed the backdoors as follows: We started with the top-level cubes (assignments) under which $R^{5}_{160}$ is satisfiable. For each such assignment $\alpha$,
we computed the backbone of $R^{5}_{160} \land \alpha$. 
The backbone can be computed using various SAT calls: Literal $l$ belongs to the backbone if and only if $R^{5}_{160} \land \alpha \land (\overline l)$ is unsatisfiable.
In most cases, the backbone assignment turned out to be a backdoor of $R^{5}_{160}$. In the remaining
cases, we extended the backbone assignments using look-aheads until they became backdoors. 
This procedure resulted in $1616$ backdoors of $R^{5}_{160}$, which---by construction---cover all satisfying assignments. 

We used these backdoors to compute all $2\,447\,113\,088$
extreme certificates $S(5, 160)$ with the {\sc sharpSAT} solver~\cite{sharpSAT} in a few seconds. 
Out of these extreme certificates, $315\,853\,824$ are modular and of the modular ones, $334\,752$ are palindromes.
The latter number has been conjectured before~\cite{Palin}.\footnote{The paper~\cite{Palin} states that the number of  palindromic extreme certificates $S(5,160)$ is $309\,408$. However, the
described method produces $334\,752$ such certificates.}





\section{Correctness}
We produced and certified a proof of unsatisfiability of $F^{5}_{161}$ to increase confidence in the correctness of our result.
The proof consists of three parts: The {\em re-encoding proof}, the {\em implication proof}, and the {\em tautology proof}. The re-encoding
proof shows the correctness of our symmetry-breaking technique, i.e., that the satisfiability of $F^{5}_{161}$ implies the satisfiability of the re-encoded formula
$R^{5}_{161}$. The implication proof includes for each cube $\alpha \in {P^5}$, a proof of unsatisfiability of 
$R^{5}_{161} \land {\alpha}$. This shows that $R^{5}_{161}$ implies each clause $C = \lnot \alpha$. 
The tautology proof shows that the disjunction of cubes is a tautology, i.e., the cubes together cover the entire search space.
Let $\overline{P^5}$ denote the negation of $P^5$, i.e., the CNF formula that has each clause $C = \lnot \alpha$ with $\alpha \in P^5$ . 
The disjunction of cubes is a tautology if and only if $\overline{P^5}$ is unsatisfiable. We show unsatisfiability of $F^{5}_{161}$ via
\[
      \mathrlap{\overbrace{\phantom{F^{5}_{161~} \models_{\!\!\!\!\mathrm{_w}} R^{5}_{161~}}}^{\text{re-encoding proof}}}
      F^{5}_{161~} \models_{\!\!\!\!\mathrm{_w}} 
      \mathrlap{\underbrace{\phantom{R^{5}_{161~} \models ~\overline{P^5}~}}_{\text{implication proof}}}
      R^{5}_{161~} \models
      \overbrace{~\overline{P^5}~ \models \bot}^{\text{tautology proof}}
\]

The proof parts have been constructed in the DRAT format, which facilitates expressing techniques that remove satisfying assignments, such as 
symmetry breaking~\cite{Heule2015}. Recent progress in verified proof checking~\cite{Cruz-FilipeMS17} reduced proof validation costs such that 
they became comparable to the costs of solving. 
We converted the DRAT proofs into LRAT proofs, a new format that was 
recently introduced to allow efficient proof checking using a theorem prover~\cite{Cruz-Filipe2017}.
We used a verified LRAT proof-checker~\cite{ITP2017}, written in the language of the ACL2 theorem
proving system~\cite{ACL2}, and applied it to certify these LRAT proofs.

Only the encoding of the Schur Number Five problem into propositional logic (i.e., the generation of $F^{5}_{161}$)  was not checked using a theorem prover. 
We chose to skip verification of this part because the encoding can be implemented using a dozen lines of straightforward C code. 

\paragraph{Re-encoding Proof.} 

The purpose of the re-encoding proof is to express our symmetry-breaking techniques---used for breaking the color symmetry $\sigma_\mathrm{col}$---in the DRAT proof system.
We did this using an existing method~\cite{Heule2015}: For each clause in a symmetry-breaking predicate, the method adds a 
new definition stating that if the lexicographical ordering is violated, two colors are swapped. However, the unit clauses $(\plit{1}{1})$ and 
$(\plit{2}{2})$ cannot be added immediately as enforcing the lexicographical order for the first two colors 
requires multiple swaps. Instead of the unit clause $(\plit{1}{1})$, we learn $(\nlit{1}{2})$, $(\nlit{1}{3})$,
$(\nlit{1}{4})$, and $(\nlit{1}{5})$ which together imply $(\plit{1}{1})$. The case of $(\plit{2}{2})$ is similar.

For example, the first unit clause in the re-encoding proof is $(\nlit{1}{5})$, which is learned as follows. 
Any assignment that assigns $\plit{1}{5}$ to true, violates the lexicographical ordering. In particular it 
violates $\plit{1}{4} \geq\plit{1}{5}$ as $\plit{1}{5}$ to true forces $\plit{1}{4}$ to false via the optional clause $(\nlit{1}{4} \lor \nlit{1}{5})$. We add constraints to the formula stating that if
$\plit{1}{5}$ is assigned to true, then every variable $\plit{i}{4}$ is swapped with $\plit{i}{5}$
and the other way around. Expressing this swap using DRAT steps requires introducing auxiliary variables. 
Afterwards the unit clause $(\nlit{1}{5})$ is implied.

The re-encoding proof is $35$ megabytes in size (uncompressed DRAT) and consists of almost a million clause addition steps and a similar number of clause deletion steps.
That is reasonably large considering that it only breaks the color symmetry $\sigma_\mathrm{col}$. However, compared to the implication proof
(discussed below) the size is negligible. 

\paragraph{Implication Proof.} 

We proved that $R^{5}_{161}$ is unsatisfiable by showing that there exists a formula, in our case $\overline{P^5}$, such that
($1$) every clause in the formula is logically implied by $R^{5}_{161}$, and ($2$) the formula can be easily shown to be unsatisfiable.
The implication proof includes, for each cube $\alpha \in P^5$, a proof of unsatisfiability of $R^{5}_{161} \land \alpha$.
The size of the implication proof is $0.88$ petabytes in the compressed DRAT format produced by {\sc Glucose} and 
$2.18$ petabytes in the compressed LRAT format produced by the {\sc DRAT-trim} proof checker. The latter format is used by 
the formally verified checker. As a comparison, the proof of the Pythagorean triples problem is 200 terabytes in the
uncompressed DRAT format~\cite{ptn}. Lightweight proof compression shrinks DRAT proofs of Schur number problems to
approximately $45\%$ of their size, while LRAT proofs are reduced to about $30\%$ of their size. DRAT proofs of Schur number
problems have lots of small numbers, while LRAT proofs have large numbers. This causes the different effectiveness in proof compression.
Based on the DRAT compression rate, the Schur Number Five proof is about ten times as large in the same format. Producing the
compressed LRAT proof required almost $20.5$ CPU years while certifying it required another $15.6$ CPU years. 

\paragraph{Tautology Proof.} 

The tautology proof describes that the disjunction of cubes is a tautology, i.e.,
that the cubes cover the entire search space. 
We showed this by proving that $\overline{P^5}$ is unsatisfiable.
The cubes produced by our partition method form a binary tree of assignments by construction. 
The tautology proof consists of 
$|\overline{P^5}|$$-$$1$ resolution steps, each time resolving two clauses whose corresponding cubes have the same parent node in the binary tree.
The size of formula $\overline{P^5}$ is $1$ gigabyte and the size of the tautology proof is $3$ gigabytes in the uncompressed DRAT format.

\subsection{Certifying the Proof}

The size of the proof demands a parallel certification approach and storing intermediate results. Below we describe our method, which
uses widely used tools.

\begin{itemize}
\item $F^{5}_{161} \models_{\!\!\!\!\mathrm{_w}} R^{5}_{161}$:
We provided the ACL2 theorem prover with the formulas $F^{5}_{161}$, $R^{5}_{161}$, and the re-encoding proof. After validating this proof,
it returns the parsed formulas $F'^{5}_{161}$ and $R'^{5}_{161}$ and a verified statement that $F'^{5}_{161} \models_{\!\!\!\!\mathrm{_w}} R'^{5}_{161}$. 
Correctness of the parsing is checked using the Unix tool {\tt diff} by comparing $F^{5}_{161}$  with $F'^{5}_{161}$ and 
$R^{5}_{161}$ with $R'^{5}_{161}$.

\item $R^{5}_{161}  \models \overline{P^5}$: We check that every clause $C \in \overline{P^5}$ is implied by $R^{5}_{161}$. The theorem prover receives
$R^{5}_{161}$, $C$, and a proof of unsatisfiability of $R^{5}_{161} \land \lnot C$.
The theorem prover returns the parsed formula $R'^{5}_{161}$, parsed clause $C'$, and a statement that $R'^{5}_{161} \models C'$. Again {\tt diff} is used to
check the equivalence of the formulas $R^{5}_{161}$ and $R'^{5}_{161}$. Clause $C'$ is stored for the next step. 

\item $\overline{P^5} \models \bot$: We construct $\overline{P'^5}$ by concatenating all clauses $C'$ implied by $R^{5}_{161}$ in the prior step, simply using the Unix tool {\tt cat}.
The theorem prover is provided with $\overline{P'^5}$ and a proof of its
unsatisfiability, and proves that the parsed formula $\overline{P''^5}$ is unsatisfiable.
The last check, again using {\tt diff}, validates that $\overline{P''^5}$ equals the stored formula $\overline{P'^5}$.
\end{itemize}


\section{Conclusions and Future Work}

We proved that $S(5) = 160$ using massively parallel SAT solving. 
To achieve this result, we designed powerful look-ahead heuristics and
developed a cheap hardness predictor to partition a hard problem into millions of manageable subproblems. These subproblems were solved
using our cube-and-conquer solver. The resulting proof is over two petabytes in size
in a compressed format. We certified the correctness of the proof using the ACL2 theorem proving system. Given the enormous size of the proof, we
argue that any result produced by SAT solvers can now be validated using highly trustworthy systems with reasonable overhead.

A century after Issai Schur proved the existence of Schur numbers, we now know the value of the first five. Determining Schur number six
will be extremely challenging and might be beyond any computational method. 
A more realistic
problem is the computation of
the fifth \emph{weak} Schur number $W\!S(5)$. 
Just a few years ago, it was shown that $W\!S(5) \geq 196$~\cite{weak}, while
it has been conjectured since the 1950s that $W\!S(5) = 196$~\cite{walker}. This appears relatively close to the value of $S(5)$.
However, we expect the corresponding propositional formula to be much harder to solve due to the lack of binary negative clauses in the encoding
of weak Schur numbers.

\section*{Acknowledgements}

The author is supported by NSF under grant CCF-1526760 and by AFRL Award FA8750-15-2-0096.
The author thanks Benjamin Kiesl, Jasmin Blanchette, Matt Kaufmann, Armin Biere, Victor Marek, Scott Aaronson, 
and the anonymous reviewers for their valuable input to improve the quality of the paper. 
The author acknowledges the Texas Advanced Computing Center (TACC) at the University of Texas at Austin for providing grid resources that have contributed
to the research results reported within this paper.

\bibliography{Schur}
\bibliographystyle{aaai}

\end{document}